\documentstyle[aps,amssymb]{revtex}		%run in artemis

\begin{document}
\draft

\title{On the Quantization of the Null-Surface Formulation of GR}
\author{
	Simonetta Frittelli$^a$\thanks{e-mail: simo@artemis.phyast.pitt.edu}
\and	Carlos N. Kozameh$^b$\thanks{e-mail: kozameh@fis.uncor.edu}
\and	Ezra T. Newman$^a$\thanks{e-mail: newman@vms.cis.pitt.edu}
\and	Carlo Rovelli$^a$\thanks{e-mail: rovelli@pitt.edu}
\and	Ranjeet S. Tate$^a$\thanks{e-mail: rstate@minerva.phyast.pitt.edu}
	}
\address{
	 ${}^a$ Department of Physics and Astronomy,
		   University of Pittsburgh,
		   Pittsburgh, PA 15260, USA				  }
\address{
	 ${}^b$ FaMAF, Universidad Nacional de C\'{o}rdoba,
		   5000 C\'{o}rdoba,
		   Argentina
        }
\date{\today}
\maketitle

\begin{abstract}

We define and discuss various quantum operators that describe the geometry
of spacetime in quantum general relativity.  These are obtained by
combining the Null-Surface Formulation of general relativity, recently
developed, with asymptotic quantization.   One of the operators defined
describes a ``fuzzy'' quantum light cone structure.  Others, denoted
``spacetime-point operators'', characterize geometrically-defined physical
points.   We  discuss the interpretation of these operators.   This
seems to suggest a picture of quantum spacetime as made of ``fuzzy''
physical points.   We derive the commutation algebra of the quantum
spacetime point operators in the linearization around flat space.

\end{abstract}

\def\comp{{\rm C}\llap{\vrule height7.1pt width1pt depth-.4pt\phantom t}}
\def\IP#1#2{\langle\, #1\, |\, #2\, \rangle}
\def\ket#1{|\, #1\, \rangle}
\def\pb#1{\rlap{\lower1ex\hbox{$\leftarrow$}}#1{}}

%-----------------------------------------------------------------

\section{Introduction}

The problem of finding and understanding the relationship between
quantum theory and gravitation is an extremely difficult
one (that has defied solution for close to seventy years) and is
simultaneously such a profound problem that it has attracted a
great deal of attention.  Its resolution could easily be a major
stepping stone to a more complete understanding of our physical
world.  The difficulties, however, are such that we might well need
radical changes in our views or completely new ideas before the
problem can be solved (See, for instance,~\cite{penrose96}).
Even if this is the case, this does not mean that we should
necessarily abandon the exploration of more traditional approaches, since
even if they fail, they could indicate possible directions to explore in
the search for the unification of gravity with quantum theory.

In this paper we present a new approach to this issue which, although
based on many of the standard ideas, differs from other approaches in
several substantial ways. In some sense our formulation lies between the
conventional and non-conventional approaches~\cite{jmp}.

The first issue we discuss is our view towards classical general
relativity (GR). At the classical level, a clear distinction can be made
between GR and other field theories. Only in GR does the geometry play a
{\em dynamical\/} role. Though often noted, this distinction has been
reemphasized in a recent series of papers by presenting GR as a theory of
characteristic {\it hypersurfaces}~\cite{fkn95a,fkn95b,fkn95c} rather than
as a theory of the metric {\em field\/}. From this point of view the
spacetime metric and associated connection are derived concepts: the basic
variables are families of 3-surfaces and a scalar function (a conformal
factor) from which a metric can be derived.  The surfaces are
automatically the characteristic surfaces of the metric and the metric
automatically satisfies the Einstein field equations. This reformulation
of GR has been referred to as the Null Surface Formulation of GR or simply
as the NSF. It appears that no other physically relevant field theory can
be stated as such a theory of surfaces.

Here, we study the quantization of the linearized version of this
approach. From this quantization of the NSF, we appear to be
led to new ideas and results on the form a quantum theory of gravity might
take. The new view essentially says that the null surfaces become
operators that obey commutation relations. Furthermore, since there is a
prescription for locating points of spacetime using foliations by families
of null surfaces, the spacetime points themselves become operators.

Roughly speaking, our formalism is a union between the Ashtekar
asymptotic quantization \cite{ashtekar87} of the gravitational field
and the NSF. In our formalism, the Bondi free data at future null
infinity ${\cal I}^+$ play a very important role. They enter as a
source in the NSF field equations. Thus, for each data set, the
solution to our classical equations represents a regular radiative
spacetime.  On the other hand, the formalism developed by Ashtekar
gives a kinematic quantization of the radiative degrees of freedom of
the gravitational field at ${\cal I}^+$.  By promoting the classical
Bondi data to quantum operators and introducing a Fock space of
asymptotic states (modulo technical difficulties addressed in detail by
Ashtekar) one is left with the ``in'' (or ``out'') states of quantum
theory. What is missing in the Ashtekar approach is the dynamical part
of the quantum theory, which would relate the asymptotic states to the
geometry of the interior of spacetime.

In this paper we adopt Ashtekar's asymptotic quantization in its
simplest form (avoiding infra-red issues) by promoting the free Bondi
data to quantum operators.  The solutions to the classical NSF
equations determine families of null surfaces in terms of this free
data. It follows that in the ``quantum theory'' the null surfaces
become operator functions of the operator data. Furthermore, since the
spacetime points are themselves determined by the intersections of the
null surfaces (and are expressible in terms of the surfaces) they can
also be thought of as operator functions of the data, with implied
non-trivial commutation relations.  We emphasize that we neither give
equal time commutation relations nor use a Hamiltonian to obtain the
``evolution'' of the operators: appropriate commutation relations for
the operator data are given on ${\cal I}^+$, and the information about
the dynamics (i.e. the full spacetime) is implicitly determined by the
NSF equations.  We emphasize that we are not discussing a field theory
on spacetime; our variables are not fields; they are surfaces composed
of spacetime points. The surfaces and associated points become the
operators. It is this feature of our point of view that appears to us
most novel.

We point out that there is no Hamiltonian for a Schr\"odinger
evolution; rather the operator ``evolution'' is given by the NSF
equations.  The formalism is most closely tied to a Heisenberg
representation.

In Section~\ref{null-surface theory} we will review some relevant
aspects of the Null Surface Formulation (NSF) of GR. (Note that we use
signature $(+,-,-,-)$.) In Section~\ref{III} we discuss what happens
when we implement the Ashtekar quantization procedure. In
Section~\ref{V} we summarize our main results and we discuss possible
meanings and ramifications of these ideas. An outline of our results
and a more detailed discussion of their physical interpretation has
appeared in \cite{fknrt96}. We relegate many of the technical details,
which can become complicated, to appendices.  In Appendix \ref{A}, as
an example, we apply our methods to the quantization of the Maxwell
theory, obtaining the standard quantization in the Coulomb gauge.

%%-----------------------------------------------------------------------

\section{Null-Surface Formulation of GR	}\label{null-surface theory}

In this section we review a new formulation, the Null Surface
Formulation of classical general
relativity~\cite{fkn95a,fkn95b,fkn95c,fkn95d}, where the emphasis has
been shifted away from more standard type of field variable
(metric, connection, holonomy, curvature, etc.) to, instead, families
of three-dimensional surfaces on a four-manifold $M^4$. [These surfaces
eventually turn out to be the characteristic surfaces of a metric.] On
the sphere bundle over $M^4$, topologically $M^4 \times S^2$, with no
further structure, there are given differential equations for the
determination of these surfaces. From the surfaces themselves, by
differentiation and algebraic manipulation, a (conformal) metric tensor
can be obtained. These surfaces, which play the role of the basic
geometric quantities, are then automatically the characteristic
surfaces of this conformal metric. Furthermore, the equations allow for
a choice of conformal factor that makes the conformal metric into a
metric which automatically satisfies the vacuum Einstein equations.  In
other words the vacuum Einstein equations are formulated as equations
for families of surfaces and a single (scalar) conformal factor.  All
geometric quantities, the metric, the connection, spin-coefficients,
Weyl and Ricci tensors, can be expressed in terms of the surfaces and
the conformal factor. In our present discussion we will be mainly
concerned with these characteristic surfaces (i.e., the conformal
structure), though of course in the full theory the conformal
factor plays an essential role.

Since the details of the differential equations are relatively
complicated~\cite{fkn95a,fkn95b,fkn95c,fkn95d} and we do not need them
for the present work, we will adopt the following strategy.  We will
assume that the differential equations for the surfaces (and conformal
factor) have been solved explicitly and then attempt to understand the
meaning of the solutions and what can be derived from them.

First of all, we have the explicit expression for the conformal
factor $\Omega= \Omega(x^a,\zeta,[data])$, where the $x^a$ are some
local pseudo-Riemannian coordinates on the manifold $M^4$, while the
$\zeta$ is a complex stereographic coordinate on the sphere, $S^2$, and
$[data]$ is the Bondi shear $[\sigma,\bar\sigma]$.  We will have little
further use here for $\Omega(x^a,\zeta)$.  Of fundamental importance
to us are the families of surfaces given as solutions to our equations,
with specific free data. They take the form
  \begin{equation}
	Z(x^a,\zeta, [data]) = u = constant.  \label{1}
  \end{equation}
{}For {\it fixed values} of $(u,\zeta)$ the above is a single function
of the four coordinates $x^a$ and thus describes a particular
three-surface. As the value of $u$ varies ({\it for fixed} $\zeta$) we
have a one-parameter foliation (of a local region) by the surfaces.
The $\zeta$ then label a sphere's worth of these foliations, i.e., a
sphere's worth of surfaces passes through each spacetime point.  {\em
Assuming\/} that the $Z$ satisfies the NSF differential equations, one
can then, in a simple and straightforward fashion, {\it obtain a
conformal metric in terms of $Z$}~\cite{fkn95a}.  Symbolically we thus
have
  \[
	g_{ab} (x^a, [data]) = g_{ab}[Z(x^a, \zeta, [data])]
  \]
where there is an overall undetermined conformal factor, essentially
$\Omega$.  Note that while $Z$ is a function of the $\zeta$, the metric
$g_{ab}$ is independent of $\zeta$.

The details of this construction are not of importance here.  What is
important is that automatically there is a (conformal) metric,
$g^{ab}(x^a)$ such that
  \begin{equation}
		g^{ab} Z,_aZ,_b = 0			\label{2}
  \end{equation}
for all $\zeta$, i.e., the surfaces $Z=constant$, are characteristic
surfaces of this metric. For simplicity, we can then choose (in a
natural fashion) a special member of the conformal class yielding an
explicit metric in terms of $Z$. (The ``naturalness'' arises from the
fact that a simple function (see below) of $Z$ is an affine parameter
for this special member of the conformal class.) We emphasize that all
conformal information about the spacetime is contained in knowledge of
$Z(x^a,\zeta)$.

For each fixed value of $\zeta$ the level surfaces of $Z$ describe a
foliation by (null) surfaces: Treating $Z$ simply as a sphere's worth
of scalar functions on $M^4$, we can construct other scalar functions
by differentiating $Z$ several times in both the $\zeta$ and
$\bar\zeta$ directions and then holding $\zeta$ constant afterwards%
\footnote{Note that differentiating $Z$ with respect to $\zeta$ is
equivalent to finding the intersections of adjacent null surfaces. For
a detailed discussion, see \cite{fkn95b}.}.  Particularly useful to us
are the two first derivatives and the mixed second derivative. Together
with the $Z(x^a,\zeta)$, these are the four functions:
 \begin{mathletters}					\label{4}
  \begin{eqnarray}
  	    u &= &             Z(x^a , \zeta, [data]), 	\label{u}    \\
       \omega & =& \eth         Z(x^a , \zeta, [data]) 	\label{om}   \\
   \bar\omega & = &\bar\eth     Z(x^a, \zeta, [data]) 	\label{barom}\\
            R &= &\eth\bar\eth Z(x^a , \zeta, [data])	\label{r}
  \end{eqnarray}
 \end{mathletters}
where $\eth$ and $\bar\eth$ are (essentially) the $\zeta$ and
$\bar\zeta$ derivatives~\cite{np66}. With the definitions
  \[
	\theta^i = (\theta^0 , \theta^+ , \theta^-, \theta^1 ) =
		   (u, \omega, \bar\omega, R)
  \]
we have
  \begin{equation}
		 \theta^i  = \theta^i (x^a, \zeta, [data]).
							\label{3.b}
  \end{equation}
These four scalar functions (parametrized by $\zeta$) have a simple
geometric meaning.
 \begin{enumerate}
    \item $\theta^0 = u = Z(x^a , \zeta, [data]) = const.$, for fixed
	$\zeta$, describes a null surface. Changing $u$ leads to a
        1-parameter foliation of $M^4$ by null surfaces.
    \item $\theta^+ = \omega = \eth Z(x^a , \zeta, [data])=const.$ and
          $\theta^- = \bar\omega = \bar\eth Z(x^a , \zeta, [data])=const.$
	chooses a null geodesic on that surface. Finally,
    \item $\theta^1= R  = \eth\bar\eth Z(x^a , \zeta, [data])$ parametrizes
	  points on that null geodesic.
 \end{enumerate}

\noindent (In fact $R$ is an affine parameter along the null geodesics for
the special member of our conformal class mentioned earlier.) The four
$\theta^i$ , for fixed $\zeta$, thus locate spacetime points.  They define
a sphere's worth of null coordinate systems, and Equation (\ref{3.b})
gives the coordinate transformation between the $\theta^i$ and $x^a$ for
each fixed $\zeta$.

Since $Z(x^a, \zeta, [data])$ contains all the conformal information of
the spacetime, so do the $\theta^i$.

An important conceptual issue is that  Eq. (\ref{3.b}) can, in
principle, be (locally) algebraically inverted into the form
  \begin{equation}
		x^a  = x^a(u, \omega, R; \zeta; [data]).
							\label{5}
  \end{equation}
Since Eq. (\ref{5}) is equivalent to (\ref{4}) it too contains the full
information about the solutions to the conformal Einstein equations;
i.e. from knowledge of Equation (\ref{5}), a metric conformal to an
Einstein metric can be obtained analytically \cite{fkn95c}. The
information about the conformal Einstein space is coded into the
functional dependence on the data.

The information about the conformal structure of spacetime, originally
encoded in $Z$ can now be extracted in an alternate manner from
(\ref{5}); a manner that is, at the moment, of direct interest to us.
If values of the $(u, \omega,\zeta)$ are chosen arbitrarily but kept
constant and $R$ is allowed to vary, Eq. (\ref{5}) is the description
of a {\it null geodesic} of the spacetime.  The five-dimensional space
of null geodesics is coordinatized by the $(u,\omega,\zeta)$, with
$(\omega,\zeta)$ complex, while $R$ parametrizes the individual
geodesics. The conformal structure is determined by the knowledge of
all null geodesics through each spacetime point, and the dependence of
these on the $[data]$ encodes the particular spacetime.  Note the dual
role equations Equations (\ref{3.b}) and (\ref{5}) play; Equation
(\ref{3.b}) describes null surfaces, its null geodesics and points on
the geodesics in terms of some ``standard" coordinates $x^a$, while
Equation (\ref{5}) describes, in parametric form, all the null
geodesics of the space. Though at first they appeared to describe the
coordinate transfomations between some null coordinates and an
arbitrary set of coordinates, $x^a$, they now have a coordinate
independent meaning.  We return to Eq. (\ref{5}) later.

%-------------------------------------------------------------------------

\subsection*{Asymptotically Flat Vacuum Spacetimes}

Before we proceed further we make the specialization from a description
of any (local) Einstein spacetime to the study of asymptotically flat
vacuum spacetimes.  In this case the geometrical meanings of the
various quantities become clearer. We begin with the fact that null
infinity, ${\cal I}^+$, exists. It can be coordinatized by a Bondi
coordinate system,
  \begin{equation}
	(u,\zeta,\bar\zeta)
  \end{equation}
with u the Bondi retarded time, and $(\zeta,\bar\zeta) \in S^2$
labeling the null generators of ${\cal I}^+$. With this notation we can
give a precise meaning to the null surfaces described by $u =
Z(x^a,\zeta, [data] )$;  they are the past null cones of the points
$(u,\zeta,\bar\zeta)$ of ${\cal I}^+$. With this meaning to $Z$ we have
a dual interpretation of $Z(x^a,\zeta)=u$, namely, if the spacetime
point $x^a$ is held constant but the $(\zeta ,\bar\zeta  )$ is varied
over $S^2$ , we obtain a two surface (topologically $S^2$) on ${\cal
I}^+$, the so-called lightcone cut of ${\cal I}^+$, defined as the
intersection of the future lightcone of the point $x^a$ with ${\cal
I}^+$.  It consists of all points of ${\cal I}^+$ reached by null
geodesics from $x^a$.  $Z$ is then refered to as the lightcone cut
function.

We have a geometric interpretation, not only of $Z(x^a,\zeta,[data])$,
but also of $\omega =\eth Z(x^a,\zeta, [data])$ and $R =\eth\bar\eth
Z(x^a,\zeta,[data])$.  $\omega$ is the ``stereographic angle'' that the
lightcone cuts make with the Bondi $u = const.$ cuts (i.e., it labels
the backward direction of the null geodesics from the point
$(u,\zeta)\in {\cal I}^+$ to $x^a$).  $R$ is a measure of the curvature
of the cut and thus a measure of the ``affine distance'' from ${\cal
I}^+$ to $x^a$ along the null geodesic.

The four functions $\theta^i(x^a,\zeta , [data])$, which are defined
geometrically on ${\cal I}^+$, describe the interior of the spacetime.
They can, be inverted (see Eq. (\ref{5})), leading to
  \begin{equation}
		x^a = x^a(\theta^i;\zeta;[data]),	\label{xtz}
  \end{equation}
which gives the location of spacetime points in terms of (geometrical)
information on ${\cal I}^+$, i.e., the $\theta^i$.

%------------------------------------------------------------------------
\subsubsection*{Linearization of the NSF}

With this (asymptotically flat) point of view, we now consider the
linearization of the Null Surface Formulation of the Einstein
equations. the coordinates used here and subsequently are the standard
cartesian coordinates $x^a$ of the background Minkowski spacetime. We
will make extensive use of this later.  In this case the conformal
factor can be taken as one; i.e.,
\begin{equation}
			\Omega(x^a, \zeta, [data]) = 1
							\label{6.a}
  \end{equation}
and the differential equation for $Z$ becomes
  \begin{equation}
	\eth^2\bar\eth^2 Z = \eth^2 \bar\sigma_{_R}(x^a,\zeta)	+
			     \bar\eth^2 \sigma_{_R}(x^a,\zeta)
			\equiv D(x^a,\zeta,[\sigma])	\label{6.b}
  \end{equation}
The data is given by a complex valued spin-weight-2 function on ${\cal
I}^+$, namely  $\sigma(u,\zeta)$ (and its complex conjugate
$\bar\sigma(u,\zeta)$) which can be given freely. The data is then
restricted to the Minkowski lightcone cut, $S^2(x^a)$ described by
(see \cite{kn83})
  \begin{equation}
   u(\zeta) = Z_0(x^a, \zeta) \equiv x^a \ell_a(\zeta),  \hspace{2cm}
		\ell_a(\zeta)\ell^a(\zeta) = 0		\label{7.a}
  \end{equation}
where  $
	\ell^a
								  =
		 \frac{1}{\sqrt{2}}
		 \big( 1					\;,\;
			   \frac{       \zeta + \bar\zeta}
				{ 1+\zeta   \bar\zeta}	\;,\;
			 -i\frac{       \zeta - \bar\zeta}
				{ 1+\zeta   \bar\zeta}	\;,\;
			   \frac{-1+\zeta   \bar\zeta}
				{ 1+\zeta   \bar\zeta}
		\big)
  $
satisfies $\eth^2 \ell^a = \bar\eth^2\ell^a = 0$ and
$Z_0(x^a,\zeta)=Z(x^a,\zeta,[0])$, i.e. $Z_0$ is the Minkowski
$Z-$function satisfying (\ref{6.b}) with zero characteristic data. Note
that $\ell^a$ and hence $Z_0$ are constructed from the first four
spherical harmonics.  Equation (\ref{7.a}), in turn, leads to the
restriction, to the lightcone cuts, of the data $\sigma(u,\zeta)$;
i.e.,
\begin{equation}
	\sigma_{_R}   (x^a, \zeta)
    =
	\sigma (Z_0(x^a, \zeta), \zeta).
							\label{7.b}
  \end{equation}
(Note that $\sigma_{_R}$ can be viewed in two different ways. It is the
pullback or restriction of $\sigma$ to a cut of ${\cal I}^+$ labelled
by the spacetime points $x^a$, but it can also be directly viewed as a
function on the sphere bundle over spacetime.) Eqs.(\ref{6.a}) and
(\ref{6.b}) are equivalent to the linearized vacuum Einstein
equations.  The general regular solution to (\ref{6.b}) is obtained as
the sum of a particular solution plus the general solution $Z_0$ to the
homogeneous equation; i.e.,
  \begin{equation}
	Z(x^a,\zeta,[data])
     =
	Z_0(x^a, \zeta)
     +
	\int_{S^2}
        	G(\zeta,\eta ) D(x^a,\eta , [\sigma ])
	dS^2_{\eta}					\label{8}
  \end{equation}
where $dS^2_\eta$ is the standard metric on the two sphere and
$G(\zeta,\eta)$ is a simple Green's function of the operator
$\eth^2\bar\eth^2$, given by
  \begin{equation}
        G(\zeta,\eta)
    =
        \frac{1}{4\pi}
                        \ell(\zeta)\!\cdot\!\ell(\eta)  \;
        \ln
           \Big(        \ell(\zeta)\!\cdot\!\ell(\eta)
           \Big)             .                           \label{Green's}
  \end{equation}
We want to point out and emphasize an important aspect of this
solution (\ref{8}). The $Z_0(x^a,\zeta)$ consists of only $l$=0,1
spherical harmonics; the second term (the particular solution) has
been chosen so that its spherical harmonic expansion contains no $l$=0,1
harmonics. One could have chosen other particular solutions {\it with}
$l=0$ or $l=1$ harmonics having as coefficients four arbitrary
functions of the  $x^a$. These four functions would constitute an
arbitrary gauge transformation in the linear theory. Our choice for
them to vanish is equivalent to a particular gauge choice. The implied
gauge is the equivalent of the Coulomb gauge of Maxwell theory,
namely, for $g^{ab}=\eta^{ab}+h^{ab}$, we have $h^{0a}=0$. The
analogous gauge choice for Maxwell theory is described in Appendix A3.

For later use, Eq. (\ref{8}) can be rewritten as
  \begin{eqnarray}
	Z(x^a,\zeta,[\sigma]) 					&
      =								&
	 Z_0(x^a, \zeta)
        +
	\int_{S^2}
		\bigg(
	   \bar\eth^2_\eta G(\zeta,\eta)\;    \sigma_{_R} (x^a,\eta)
	+      \eth^2_\eta G(\zeta,\eta)\;\bar\sigma_{_R} (x^a,\eta)
		\bigg)
	d^2\!S_\eta					\nonumber\\
								&
     \equiv							&
	 Z_0(x^a, \zeta)
        +
	 Z_1(x^a, \zeta,[\sigma])			\label{ultimatez}
  \end{eqnarray}
This expression is obtained from Equation (\ref{8}) by using properties
of the Green's function $G(\zeta,\eta)$ (see
Appendix~\ref{section:01Green's}), and from the assumption that
$\sigma_{_R}$ is a regular function on the sphere. By differentiation
(with respect to $\zeta$) of (\ref{8}) it is a simple matter to
construct the full set of $\theta^i$, i.e. Eqs.~(\ref{u}-\ref{r}), and
invert them explicitly to obtain Equations~(\ref{5}).  The explicit
linearized inversion is given in Section~\ref{qspt}.

Simply for completeness, we mention that the full set of Einstein
equations are a generalization of Equations (\ref{6.a}) and
(\ref{6.b}); Equation (\ref{6.a}) for the conformal factor becomes more
complicated, while (\ref{6.b}), the equation for $Z$, retains the same
form; it has an additional, rather complicated term added to the
right-hand side that does depend on the $\Omega$.

%------------------------------------------------------------------
\section{Quantization of linearized GR}		\label{III}

In the previous section we described how the classical data on ${\cal
I}^+$ can be used to reconstruct various geometrical structures in the
interior of the spacetime: null surfaces, null geodesics and the
locations of spacetime points in a given local chart.  In this section,
by analogy, we begin with an asymptotic quantum theory at ${\cal
I}^+$~\cite{ashtekar87}, and subsequently extend it into the interior
of the spacetime.  We implement this idea by constructing quantum
operators corresponding to the various geometrical entities described
in the previous section.  We finally compute various physically
interesting commutation relations obtained from the free-field
commutation relations on the data at ${\cal I}^+$.

While most of our calculations are formal, all quantities (in the
linearized case) can be defined rigorously on the asymptotic Fock
space.  Alternatively, we can think of all quantities as abstract
operators subject to non-trivial commutation relations.

In the first subsection we briefly introduce the asymptotic quantum
theory (done in detail for the free Maxwell field in Appendix A1),
essentially the quantization of the characteristic free data at ${\cal
I}^+$, and describe the construction of the asymptotic Fock space (the
details are given in Appendix A2).  We describe Ashtekar's asymptotic
quantization~\cite{ashtekar87}, differing only in notational details.
In addition, we ignore infra-red sectors.

The remaining subsections contain the construction of the new quantum
operators.  Since all of them have a functional dependence on the data
through the $Z$ function, our first result is the quantization of the
null surfaces, in subsection~\ref{quantumz}.  The commutator for the
$Z$ function at two different points is, then, of fundamental
importance to the remainder of the section, in which we construct the
quantum analogs of the various geometrical quantities (subsection
\ref{qcuts}) and quantum spacetime points (subsection \ref{qspt}).

\subsection{Asymptotic quantum theory}

As is quite well-known, the radiative degrees of freedom of the
gravitational field are specified by the characteristic initial data on
${\cal I}^+$. The space of characteristic initial data is a phase space
coordinatized by either the Bondi shear
$\sigma_{ab}(u,\zeta)=\sigma(u,\zeta)\bar{m}_a\bar{m}_b+
\bar\sigma(u,\zeta)m_am_b$, or the complex Bondi news
$N_{ab}=N(u,\zeta)\bar{m}_a\bar{m}_b+\bar{N}(u,\zeta)m_am_b$. The
complex Bondi shear $\sigma(u,\zeta)$ serves as a potential for the
complex Bondi news $N=\partial \sigma/\partial u$. The action of the
symplectic form on two vectors (infinitesimal news scalars) $\delta
N_1$ and $\delta N_2$ tangent to the phase space of characteristic data
is~\cite{ashtekar87}
  \begin{equation}\label{sympgr}
	\Omega(\delta N_1,\delta N_2)
     =
	\frac1{2\pi} 	\int
			\int_{{\cal I}^+}\> du\, dS^2\,\>
						du'\, dS^2\,'\>
	\delta^2(\zeta-\zeta')\Delta(u-u')\>
	\left(\delta N_1(u,\zeta)
   \bar{\delta N}_2(u',\zeta')-\delta N_2(u,\zeta)
   \bar{\delta N}_1(u',\zeta')\right),
  \end{equation}
where $\Delta(u)=\frac12 \mbox{sgn}(u)$ is the skew-symmetric
anti-derivative of $\delta(u)$, so that
$\delta(u)=\partial\Delta(u)/\partial u$ (as distributions); and
$du\,dS^2=-2i du\wedge \,d\zeta\wedge d\bar\zeta/(1+\zeta\bar\zeta)^2$
is the volume element on ${\cal I}^+$.  Note that this phase space is
analogous to the phase space for source-free Maxwell theory, with
$\sigma,N$ playing the roles of $A,E$ respectively.  Thus the
asymptotic aspects of the quantization are identical to the
construction detailed in Appendix \ref{A}.

Consider the space ${\cal S}$ of $C^\infty$ spin-weight-2 complex
scalar fields $N$ on ${\cal I}^+$, all of whose components in a
$(u,\zeta)$ chart and all their derivatives fall-off faster than
$1/|u|^n$ for any $n$, for large values of $|u|$.  On the positive
frequency (with respect to $u$) subspace ${\cal S}^+$ of news
functions, one can introduce a Hermitian inner product analogous to
(\ref{ip1}). One can then Cauchy complete this space to obtain the one
particle Hilbert space, on which one constructs the asymptotic Fock
space of the characteristic data for the radiative modes of GR. In a
fashion analogous to that for the free Maxwell field, one then
constructs operator-valued distributions corresponding to the Bondi
news, and the Bondi shear $\sigma(u,\zeta)$. These operator-valued
distributions satisfy~\cite{ashtekar87} the formal commutation
relations given by
  \begin{equation}
	[\widehat{    \sigma}(u ,\zeta ),
	 \widehat{\bar\sigma}(u',\zeta')]
      =
	-2\pi i\hbar\Delta(u-u')\delta^2(\zeta-\zeta')\hat1,
						\label{freccr}
  \end{equation}
where $\delta^2(\zeta-\zeta')$ has spin weight 2 in $\zeta$ and $-2$ in
$\zeta'$, and is defined such that
$\int_{S^2}\delta^2(\zeta-\zeta')f(\zeta')dS'^2=f(\zeta)$ for all spin
weight $+2$ functions $f$ (see \cite{np66} for the treatment of
$\delta$ functions in the context of spin-$s$ spherical harmonics).
These are the fundamental commutation relations for the data on ${\cal
I}^+$. Since all the other operators are constructed via their
functional dependence on the data, these commutation relations are
critical to obtaining the commutation relations between the interesting
geometrical operators.

%------------------------------------------------------------------
\subsection{Quantum hypersurfaces} \label{quantumz}

We now present a construction that extends the quantization available
at ${\cal I}^+$ into the interior of the spacetime.  In a rather
non-standard fashion, we proceed to the quantization of hypersurfaces and
spacetime points, instead of the more traditional approach of
quantizing the metric fields or connections.  This construction is
based on the null-surface formulation of GR and a (classical)
{\em dynamical prescription\/} to specify a location in the interior
manifold.  In Section \ref{null-surface theory}, we had two dynamical
prescriptions, with different meanings: Equation (\ref{3.b}),
$\theta^i=\theta^i(x^a, \zeta, [\sigma])$, which for given $x^a, \zeta$
and $\sigma$ define four null-geodesic quantities; or Equation
(\ref{5}), $x^a=x^a(\theta^i, \zeta, [\sigma])$, which for given values
of $\theta^i$ and $\zeta$ (fixed $\sigma$) locates an interior
spacetime point.

Both alternatives require the explicit expression for the function
$Z(x^a, \zeta,[\sigma])$, obtained in Section 2 (\ref{ultimatez}). $Z$
can be viewed as describing null hypersurfaces of the spacetime by
setting $Z(x^a, \zeta,[\sigma])=const.$ Therefore, we will first
develop the formal quantization of $Z(x^a,\zeta,[\sigma])$, without
attempting to give it a meaning immediately.

We define the operator $\widehat{Z}$ by simple substitution, in
(\ref{ultimatez}), of the classical variables $\sigma$ with their
quantum analogs $\widehat{\sigma}$; i.e.,
  \begin{eqnarray}
	\widehat{Z}(x^a, \zeta) 				&
     \equiv 							&
		  Z(x^a, \zeta,      [\widehat{\sigma}])	\nonumber\\
								&
    = 								&
		  Z_0(x^a, \zeta) \;\; \widehat{1}
      +
	\int_{S^2}
		\bigg(
        \bar\eth^2_\eta                            G(\zeta,\eta)\;
			\widehat{    \sigma}(Z_0(x^a,\eta),\eta)
      +
	    \eth^2_\eta	                           G(\zeta,\eta)\;
			\widehat{\bar\sigma}(Z_0(x^a,\eta),\eta)
		\bigg)\;\,
								\label{zhat}
	d^2\!S_\eta
  \end{eqnarray}
The operator $\widehat Z$ is manifestly linear in the free data
$\widehat \sigma$. The free-data commutation relations (\ref{freccr})
imply the following integral representation of the commutation
relations for $\widehat Z$
  \begin{eqnarray}
		[ 	\widehat{Z}		,
			\widehat{Z}'
		  ]						&
    \equiv							&
		[	\widehat{Z}(x^a, \zeta )	,
			\widehat{Z}(x'^a,\zeta')
		  ]					\nonumber\\
								&
    =								&
	-2\pi i\hbar	\!
	\int_{S^2}	\!\!
	   \bigg(
		\bar\eth^2_\eta G(\zeta ,\eta)
		    \eth^2_\eta G(\zeta',\eta)
	+
		    \eth^2_\eta G(\zeta ,\eta)
		\bar\eth^2_\eta G(\zeta',\eta)
	   \bigg)
		\Delta(y\!\cdot\!\ell(\eta))	\;
	d^2\!S_\eta				\;\;\;
			\widehat{1}			\label{intzccr}
  \end{eqnarray}
where we use the notation
  $
		v\!\cdot\!w
       \equiv
		v^a\eta_{ab}w^b
  $
for vectors $v^a$ and $w^a$ on Minkowski space, $y^a\equiv x^a-x'^a$
and the vector $\ell^a$ was introduced in equation (\ref{7.a}).

The commutator $[\widehat{Z},\widehat{Z}']$ is symmetric under
interchange of only $\zeta$ with $\zeta'$ and antisymmetric under
interchange of only $x^a$ with $x'^a$. The latter antisymmetry implies
that $[\widehat{Z},\widehat{Z}]$ and its $\zeta$ derivatives vanish
identically, a property that has important consequences in the
following two subsections.

The evaluation of the commutator (\ref{intzccr}) in closed form is a
cumbersome calculation.  In the case of {\bf timelike} $y^a$ the
closed-form commutator is:
  \begin{equation}
	 [\widehat{Z},\widehat{Z}']
   =
	-2\pi i\hbar
		\Big(
				\ell\!\cdot\!\ell'
			\ln (	\ell\!\cdot\!\ell'  )
		      + \frac13
		      - \frac16	\ell\!\cdot\!\ell'
		\Big)
			\Delta(x^0\!\!-\! x'^0)	\;\;
	  \widehat{1}					\label{closedzccr}
  \end{equation}
where $x^0$ and $x'^0$ are the time components of $x^a$ and $x'^a$
respectively. The calculation follows essentially the same steps as in
the analogous case of Maxwell fields, which we include in
Appendix~\ref{section:closedccr}. This calculation is considerably
simpler than the case of spacelike separation because, in the timelike
case, the step function $\Delta(y\!\cdot\!\ell)$ takes a constant value
on the sphere (+$\frac12$ if $y^a$ is future pointing, or $-\frac12$ if
$y^a$ is past pointing).  If the step function changes sign on the
sphere, as in the spacelike case, there is a non-vanishing line
integration on the boundary where the sign change takes place.  This
line integral becomes lengthy and cumbersome (though straightforward)
to evaluate (see Appendix~\ref{section:closedccr} for a very similar
calculation in the case of Maxwell fields). Though this calculation has
not yet been completed, it is not clear that the closed form will
shed light on the discussion that follows.

In the remainder of this section, we turn our attention to the
interpretations of two of the several alternate quantum descriptions
which arise from the fact that $Z$ is quantized.

%------------------------------------------------------------------
\subsection{Quantum lightcone cuts and associated geometric quantities}
							\label{qcuts}

Consider Eqs. (\ref{4}).  At the classical level, they define four
geometric quantities associated with null surfaces (see Section
\ref{null-surface theory}).  In the linearization, they are explicitly
given by
  \begin{eqnarray}
	u							&
    =								&
		          Z_0   + Z_1
   =
	  x^a\ell_a             + Z_1(x^a,\zeta,[\sigma])		\\
	\omega							&
    =								&
                     \eth Z_0   + \eth Z_1
   =
	  x^a   m_a         + \eth Z_1(x^a,\zeta,[\sigma])		\\
     \bar\omega							&
    =								&
                 \bar\eth Z_0   + \bar\eth Z_1
    =
	  x^a\bar{m}_a     + \bar\eth Z_1(x^a,\zeta,[\sigma])	\\
	R							&
    =								&
             \eth\bar\eth Z_0   + \eth\bar\eth Z_1
    =
	  x^a(n_a-\ell_a)       + \eth\bar\eth Z_1(x^a,\zeta,[\sigma])
							\label{tetrad}
  \end{eqnarray}
where $m_a\equiv\eth\ell_a$, $\bar{m}_a\equiv\bar\eth l_a$ and
$n_a\equiv\eth\bar\eth\ell_a + \ell_a$,
or
  \begin{equation}
	\theta^i
     =
	x^a\lambda^i_a(\zeta)	+
	\theta^i_1    (x^a,\zeta,[\sigma])		\label{firsth}
  \end{equation}
where $\theta^i_1(x^a,\zeta,[\sigma])\equiv(Z_1,\eth Z_1,\bar\eth
Z_1,\eth\bar\eth Z_1)$ and $\lambda^i_a(\zeta) \equiv
(\ell_a,m_a,\bar{m}_a,n_a-\ell_a)$.  For future reference we recall
that the four vectors $\ell_a, m_a, \bar{m}_a$ and $n_a$ satisfy
$\ell_a n^a=-m_a\bar{m}^a=1$, while the remaining scalar products
among any two of them are zero.  Furthermore, $n^a + \ell^a =
\sqrt{2}\delta^a_0$.

We now define a set of quantum operators, in the following way
  \begin{equation}
		\widehat{\theta}^i(x^a,\zeta)
   \equiv
	 	\theta^i(x^a,\zeta,[\widehat{\sigma}])	\;\;.
  \end{equation}
Explicit expressions of these in terms of the data can be obtained from
(\ref{firsth}) and (\ref{ultimatez}), namely
  \begin{eqnarray}
    				   \widehat{u         } 		&
   \equiv 	   							&
		x^a \ell_a(\zeta)\;\widehat{1         }		\;+\;
		   Z_1(x^a,\zeta, [\widehat{\sigma    }]),	\label{tuno}\\
    				   \widehat{\omega    }			&
   \equiv     								&
	       x^a     m_a(\zeta)\;\widehat{1         }		\;+\;
              \eth Z_1(x^a,\zeta, [\widehat{\sigma    }]) \label{tdos}	\\
    				   \widehat{\bar\omega} 		&
   \equiv 								&
               x^a\bar{m}_a(\zeta)\;\widehat{1         } 	\;+\;
	  \bar\eth Z_1(x^a,\zeta, [\widehat{\sigma    }]) 		\\
	    			\widehat{R         }			&
   \equiv								&
        x^a (n_a(\zeta)-\ell_a(\zeta))\;\widehat{1         }		\;+\;
      \eth\bar\eth Z_1(x^a,\zeta, [\widehat{\sigma    }])	\;\; .
\label{tquattro}
  \end{eqnarray}
They are manifestly linear in $\widehat{\sigma}$.

$\widehat{\theta}^i(x^a,\zeta)$ constitute a set of four quantum
operators depending on $(x^a,\zeta)$.  Therefore, in this picture, the
interior points $x^a$ are considered as $c$-numbers, whereas
$\widehat{\theta}^i$, the geometric structures at ${\cal I}^+$, are
quantum variables, subject to possible fluctuations.

The commutator $[\widehat{u},\widehat{u}'] \equiv
[\widehat{u}(x^a,\zeta),\widehat{u}(x'^a,\zeta')]$ is simply
$[\widehat{Z},\widehat{Z}']$, obtained earlier; i.e., Equation
(\ref{intzccr}).  The other commutators
$[\widehat{\theta}^i,\widehat{\theta}'^j] \equiv
[\widehat{\theta}^i(x^a,\zeta),\widehat{\theta}^j(x'^a,\zeta')]$ can be
obtained by differentiation of $[\widehat{Z},\widehat{Z}']$:
%%
%%	In the following, the extra braces are necessary since
%%	latex mistakes [] for an optional argument, though
%%	not in all cases...
%%
 \begin{equation}
 \begin{array}{l}
			[ \widehat{u     },\widehat{u}'          ]
    =		        [ \widehat{Z     },\widehat{Z}'          ]     	\\
		       {[ \widehat{u     },\widehat{\omega}'     ]}
    =		\eth'   [ \widehat{Z     },\widehat{Z}'          ]     	\\
		       {[ \widehat{u     },\widehat{\bar\omega}' ]}
    =       \bar\eth'   [ \widehat{Z     },\widehat{Z}'	         ]	\\
		       {[ \widehat{u     },\widehat{R}' 	 ]}
    =  \eth'\bar\eth'	[ \widehat{Z     },\widehat{Z}'          ]    	\\
		       {[ \widehat{\omega},\widehat{\omega}'     ]}
    =		\eth
                \eth'	[ \widehat{Z     },\widehat{Z}'          ]
  \end{array}
  \hspace{1.5cm}
  \begin{array}{l}
		       {[ \widehat{\omega    },\widehat{\bar\omega}' ]}
    =		\eth
            \bar\eth'	[ \widehat{Z         },\widehat{Z}'          ]	\\
		       {[ \widehat{\omega    },\widehat{R}'          ]}
    =		\eth
       \eth'\bar\eth'	[ \widehat{Z         },\widehat{Z}'          ]  \\
		       {[ \widehat{\bar\omega},\widehat{\bar\omega}' ]}
    =       \bar\eth
    	    \bar\eth'   [ \widehat{Z         },\widehat{Z}'          ]  \\
		       {[ \widehat{\bar\omega},\widehat{R}'          ]}
    =       \bar\eth
       \eth'\bar\eth'   [ \widehat{Z         },\widehat{Z}'          ]	\\
		       {[ \widehat{R         },\widehat{R}'          ]}
    =  \eth \bar\eth
       \eth'\bar\eth'	[ \widehat{Z         },\widehat{Z}'          ]
 \end{array}
							\label{thetaccr}
 \end{equation}
It can be inferred from (\ref{intzccr}) that these commutators are,
generically, non-vanishing functions of $x^a$, $x'^a$, $\zeta$ and
$\zeta'$ (the closed forms are lengthy and complicated).  The immediate
consequence of the non-vanishing of the commutators is that the four
geometric operators $\theta^i$ do not have a complete set of common
eigenstates.
Furthermore, since a generic state is not an eigenstate of any of the
four operators, in a generic state, all four geometric quantities
will fail to have
well-defined values.  In this sense, the lightcone cut ($u$), its
curvature ($R$) and the angle of emittance ($\omega$) of the null
geodesics at ${\cal I}^+$ are ``fuzzy''.
%An interesting issue here is
%whether there is a subset of the parameters $(x^a,x'^a,\zeta,\zeta')$
%such that the $\widehat{\theta}^i$ commute. The significance of an
%affirmative answer to this question remains to be clarified.

%-----------------------------------------------------------------------

\subsection{Quantum spacetime points}		\label{qspt}

We now consider the ``dual'' picture, which arises from the inversion
(\ref{5}).  Classically, the $x^a$ represent an interior spacetime point
which can be reached from ${\cal I}^+$ by specifying the values of:
{\it i)} the observation point $(u,\zeta)$ at ${\cal I}^+$, {\it ii)}
the angle $\omega$ of the null geodesic emitted inwardly from
$(u,\zeta)$, aimed at $x^a$, and {\it iii)} the focusing distance $R$
along the null geodesic $(u,\zeta,\omega)$ at which the point $x^a$ is
located. The linearized version of (\ref{5}) can be obtained from
(\ref{firsth}) in the form
  \begin{equation}
	x^a(\theta^k,\zeta,[\sigma])
     =
	\lambda^a_i(\zeta)\;
	\theta^i
      -
	\lambda^a_i(\zeta)\;
	\theta^i_1 (\lambda^a_j(\zeta)\theta^j,\zeta,[\sigma])	\;\;,
  \end{equation}
where by $\lambda^a_i(\zeta)$ we denote the inverse matrix to
$\lambda^i_a(\zeta)$, namely
$\lambda^a_j(\zeta)\lambda_a^i(\zeta)=\delta^i_j$, explicitly given
by
  \begin {equation}
	\lambda^a_i(\zeta)
     =
      (	\lambda^a_0,
	\lambda^a_+,
	\lambda^a_-,
	\lambda^a_1
	)
    =
			(	               n^a \!+\! \ell^a,
			       -         \bar{m}^a	       ,
			       -               m^a	       ,
					    \ell^a
				)				\;\;.
  \end{equation}

We now define the operators associated with the spacetime points
$x^a$ as
  \begin{equation}
		\widehat{x}^a(\theta^i,\zeta)
      \equiv
			  x^a(\theta^i,\zeta,[\widehat{\sigma}])
      =
		\lambda^a_i(\zeta)\;
	\theta^i		  \;	\widehat{1}
      -
	\lambda^a_i(\zeta)\;
	\theta^i_1 (\lambda^a_j(\zeta)\theta^j,\zeta,[\widehat{\sigma}])
\label{tcinque}
  \end{equation}
and obtain a quantized description of the interior spacetime points
$x^a$. Now the surface quantities $\theta^i$ remain $c$-numbers.
$\widehat{x}^a(\theta^i,\zeta)$ constitute a set of four operators
dependent on the six parameters $(\theta^i,\zeta)$.

Since the spacetime-point operators $\widehat{x}^a$ are functions of
the fundamental operators $\widehat{\sigma}$, they also are subject to
commutation relations $[\widehat{x}^a,\widehat{x}'^b]\equiv
[\widehat{x}^a(\theta^i,\zeta),\widehat{x}^b(\theta'^i,\zeta')]$ which
can be derived from $[\widehat{\sigma},\widehat{\bar\sigma}']$. The
commutators are
  \begin{equation}
			       [\widehat{x}^a			   ,
				\widehat{x}'^b			    ]
	=
		\lambda^a_i
		\lambda'^b_j				\;\;
			       [\theta^i_1
	       (\lambda^c_k     \theta^k ,\zeta ,[\widehat{\sigma}] ),
				\theta^j_1
	       (\lambda'^c_k    \theta'^k,\zeta',[\widehat{\sigma}']) ]	\;,
							\label{xccr}
  \end{equation}
where the commutators
  $
			       [\theta^i_1
	       (\lambda^c_k     \theta^k ,\zeta ,[\widehat{\sigma}] ),
				\theta^j_1
	       (\lambda'^c_k    \theta'^k,\zeta',[\widehat{\sigma}']) ]
  $
are found from (\ref{intzccr}) by using (\ref{thetaccr}). Explicitly,
\begin{eqnarray}
				[\widehat{x}^a,\widehat{x}'^b ]		&
      =									&
		        2
        \delta^a_0
	\delta^b_0
				[\widehat{Z},\widehat{Z}']
      +	          \sqrt{2}
       	\delta^a_0
	 \ell'^b
		  \eth'\bar\eth'[\widehat{Z},\widehat{Z}']
      -	          \sqrt{2}
      	\delta^a_0
      \bar{m}'^b
		           \eth'[\widehat{Z},\widehat{Z}']
      -	          \sqrt{2}
      	\delta^a_0
            m'^b
		       \bar\eth'[\widehat{Z},\widehat{Z}']	\nonumber\\
									&
									&
      +	          \sqrt{2}
	\ell^a
      	\delta^b_0
		   \eth\bar\eth	[\widehat{Z},\widehat{Z}']
      +	\ell^a
	\ell'^b
		  \eth\bar\eth
		  \eth'\bar\eth'[\widehat{Z},\widehat{Z}']
      -	\ell^a
    \bar{m}'^b
		  \eth\bar\eth
		           \eth'[\widehat{Z},\widehat{Z}']
      -	\ell^a
          m'^b
		  \eth\bar\eth
		       \bar\eth'[\widehat{Z},\widehat{Z}']	\nonumber\\
									&
									&
     -	          \sqrt{2}
       \bar{m}^a
	\delta^b_0
		   	   \eth
				[\widehat{Z},\widehat{Z}']
     - \bar{m}^a
	 \ell'^b
		  \eth
		  \eth'\bar\eth'[\widehat{Z},\widehat{Z}']
       +  \bar{m}^a
         \bar{m}'^b
		  	   \eth
		           \eth'[\widehat{Z},\widehat{Z}']
      +	\bar{m}^a
             m'^b
		           \eth
		       \bar\eth'[\widehat{Z},\widehat{Z}']	\nonumber\\
									&
									&
     -	          \sqrt{2}
	     m^a
	\delta^b_0
		       \bar\eth
				[\widehat{Z},\widehat{Z}']
      -	    m^a
	\ell'^b
		  \bar\eth
		  \eth'\bar\eth'[\widehat{Z},\widehat{Z}']
       +        m^a
         \bar{m}'^b
		       \bar\eth
		           \eth'[\widehat{Z},\widehat{Z}']
      +	        m^a
               m'^b
		       \bar\eth
		       \bar\eth'[\widehat{Z},\widehat{Z}'],
							\label{explccr}
 \end{eqnarray}
where $[\widehat{Z},\widehat{Z}']$ is given by (\ref{intzccr})
evaluated at $x^a=\lambda^a_k\theta^k =u(n^a+\ell^a) +R \ell^a
-\omega\bar{m}^a -\bar{\omega}m^a$ and $x'^a=\lambda'^a_k\theta^k
=u'(n'^a+\ell'^a) +R' \ell'^a -\omega'\bar{m}'^a -\bar{\omega}'m'^a$.
We have thus obtained non-trivial commutators for operators which
correspond to the coordinates of spacetime points.  A series of
conceptual issues arise from the existence of the non-trivial
commutators.  In this quantum picture, we would like to define the
notion of spacetime point. Classically, a spacetime point can be
specified by giving a $4-$tuple of numbers, the values of the
coordinates $x^a$ on a $4-$manifold.  In the quantum description,
however, an operator $\widehat{x}^a$ (fixed $a$) takes a well-defined
value only when acting on an eigenstate and a set of operators (all
$\widehat{x}^a$) have a complete set of simultaneous eigenstates if and
only if all pairs mutually commute. Let us explore what kind of an
analog of a spacetime point we can construct.

Let us fix the values of the classical parameters $\theta^i;\zeta$.
Classically, these define the spacetime point whose coordinates are
$x_{cl}^a=x^a(\theta^i;\zeta)$ (see (\ref{5})). An important question
at this juncture is whether the set of four operators
$\widehat{x}^a(\theta^i;\zeta)$ form a commuting set. It can be checked
by inspection, setting $\theta'^k=\theta^k$ and $\zeta'=\zeta$ in
(\ref{explccr}), that all four operators
$\widehat{x}^a(\theta^i;\zeta)$ do commute with each other, as a
consequence of the vanishing of $[\widehat{Z},\widehat{Z}]$ and all its
$\zeta-$derivatives (see the discussion after (\ref{intzccr}).
Therefore we can define the quantum analog of a spacetime point as a
common eigenstate of the four coordinates $\hat{x}^a$.  Let us denote
this eigenstate by $|x^a_{\theta^i;\zeta}\rangle$. Now note that the
eigenvalues of the operators $\widehat{x}^a(\theta^i;\zeta)$, which are
denoted by $x^a_{\theta^i;\zeta}$, can in general take a wide range of
values and need not be equal to $x^a_{cl}$. Thus, in any state of
quantum gravity, there is a ``probability of finding'' the spacetime
point defined by $(\theta^i;\zeta)$ at values other than the classical
value $x^a_{cl}=x^a(\theta^i;\zeta)$.

Next let us consider whether all spacetime points can simultaneously be
assigned values. This would require that the right hand side of
(\ref{explccr}) vanish identically. However, if $\theta'^k \neq
\theta^k$ and $\zeta' \neq \zeta$ the commutator between two separate
spacetime-point operators $\widehat{x}^a$ and $\widehat{x}'^b$ is
generically non-vanishing. Thus there are no common eigenstates of all
the distinct spacetime points, and as a consequence, we have no
candidate for a quantum analog of the spacetime manifold. Another way
to see this is that in a common eigenstate of a particular
spacetime-point set of operators, only that one point in the manifold
is well-defined, while the rest of the manifold becomes ``fuzzed''
out.  In our second quantum picture, then, the interior spacetime is
lost as a distinct classical manifold.

On the technical side, the commutators (\ref{explccr}) display a
singular behavior at the points $\zeta = \zeta'$, which makes the
exploration of the ideas in the preceding paragraph a complicated
task.  Removal of the $S^2$'s degrees of freedom from the commutators
has been tried by means of double integration on the sphere, with the
unsatisfactory result that the commutators (\ref{explccr}) {\em
vanish\/} upon integration.

%-------------------------------------------------------------------------

\section{Remarks}			\label{V}

In this final section we will summarize our results
and discuss their relevance to the issues of quantum spacetime.

By combining Ashtekar's asymptotic quantization of the gravitational
field with the Null Surface Formulation of GR we have (in the linear
version) constructed certain non-standard operators on the quantum
state-space. The classical variables (to which these operators
correspond) are not, in any conventional sense, the usual or standard
field variables: they are families of point sets, specifically,
families of three-dimensional surfaces. Though the surfaces are
described by functions, it is the surfaces themselves which are
fundamentally important, not the numerical values associated with them.
Thus, it is not important if the functions that describe the surfaces
are ``large'' or ``small'' or even whether they ``vanish''. From
knowledge of these surfaces, all null geodesics, light-cones and the
conformal structure of a space-time can be constructed.  By analyzing
the intersections of these surfaces one could even ``pick out'' or
choose space-time points \cite{fkn95b}.  It is possible to even think
of these surfaces as being the primitive elements of the theory with
the space-time points and light-cones as derived concepts.  One thus
sees that the associated operators are not, in any obvious fashion,
standard field operators.  Instead, we have operators that correspond
to null surfaces, null geodesics and field ``points''.  The novelty of
this approach to quantum gravity lies in this feature.  It appears to
be saying that it is the spacetime itself, i.e., the manifold
structure, that is undergoing the quantization process and not, as in
the more standard approaches, some metric or connection field.

More specifically, the first and most important of our
operators is $\widehat{Z}(x^a,\zeta)$, defined in (\ref{zhat}). The
classical analog $Z(x^a,\zeta)$ determines the characteristic surfaces
in the NSF.  In the ``presumed'' quantum theory, only the average
position of the surfacen (whatever interpretation one might give to
that) is determined for any given quantum state, by the expectation
value of the operator. The ``observed'' position can be predicted only
probabilistically.

The other operators of the set $\widehat\theta^i$, i.e.,
Eq.(\ref{tdos}-\ref{tquattro}), for asymptotically flat spaces,
correspond to simple classical geometric objects, angles at ${\cal
I}^+$ labelling null geodesics (directions of sight) and curvatures of
light-cone cuts (focus distances) at ${\cal I}^+$. Once again, as
quantum operators they are non-conventional; nevertheless ``observed''
values are probabilistically determined.

The third, and perhaps most interesting, family of operators is given
by the ``spacetime point'' operators $\widehat{x}^a(\theta^i;\zeta)$,
defined in (\ref{tcinque}).  Let us discuss an aspect of their
classical physical meaning.  In order to fix ideas physically, imagine
that we wish to describe a gravitational phenomenon localized in a
certain spacetime region $\cal R$, which we consider to be small.
Consider the classical quantities ${x}^a(\theta^i;\zeta) =
{x}^a(u,\omega,R;\zeta;[data])$.  The three independent variables
$u,\zeta$ determine a point on future (null) infinity ${\cal I}^+$.
Recall that $\zeta$ coordinatizes the celestial sphere, and $u$ the
Bondi time.  One may think of $u,\zeta$ as labeling asymptotic
observers.  Imagine that these observers look into the region $\cal
R$.  Each of them can vary the direction of sight, labelled by the
independent variable $\omega$.  Finally, using a focussing distance
labelled by the variable $R$, each of them can determine the distance
to a point in ${\cal R}$.  Thus, the set  $(u,\omega,R;\zeta)$
determines the locations of observers and the direction of sight and
focus distance of their observations, looking into $\cal R$ from a
surrounding region.  Now, since the trajectories of light rays are
determined by the gravitational field, the actual point $x^a$ seen by
the observer at $(u,\zeta)$ looking at a distance $R$ in the direction
$\omega$ depends on the gravitational field. For a given spacetime, the
quantities ${x}^a(u,\omega,R;\zeta;[data])$ determine this point.

It is a rather remarkable fact that these quantities, $x^a =
x^a(u,\omega,R;\zeta; [data])$, specify the conformal space-time
geometry uniquely. Let us describe them in slightly more detail before
returning to the quantum case.  Consider the 6-dimensional
``observation space'' defined by the three coordinates $(u,\zeta)$ of
an observer's position on ${\cal I}^+$, the two angles of observation,
$\omega$, and the focus distance $R$.  On this observation space
consider a four parameter family of two dimensional surfaces,
topologically $S^2$ -- each two-surface will be referred to as a leaf
and the leaves foliate the observation space. Our equations $x^a =
x^a(u,\omega,R;\zeta)$ are precisely of this form, i.e., each spacetime
point $x^a$ is equivalent to a leaf.  (Notice furthermore, that it is
the family of leaves that defines the space-time points geometrically
even if we change the gauge arbitrarily to $y^a = f^a(x^b)$.)
Physically, this amounts to saying that a spacetime point can be viewed
as the collection of points in observation space, i.e., (locations,
directions of sight, focus-distances) from which surrounding observers
see it.  Remarkably, this foliation by the equivalence classes of
points in the observation space that ``see'' the same space-time point
is equivalent to giving the conformal pseudo-Riemannian
geometry~\cite{fkn95d}.

In the quantum domain, it is worth asking what validity this picture
might have even when the spacetime geometry undergoes ``quantum
fluctuations''. The equations that define the leaves become operator
equations, i.e. $\widehat{x}^a = x^a(u,\omega,R;\zeta;
\widehat{[data]})$.  Now imagine that we are in the realm of quantum
gravity. Then it is difficult to imagine how we could identify points
physically inside ${\cal R}$.  However the construction partially
survives.  The ``observation space'' remains classical and hence we
still have a family of observers surrounding ${\cal R}$ and looking in;
specifically, the observers' locations, their directions of sight and
focus distances are still labelled by the classical parameters
$(u,\omega,R;\zeta)$.  What changes is that for a fixed quantum state,
we will not have a sharply defined value for the operator
$\widehat{x}^a$ (the leaf) - except when it is in an eigenstate - but
only a probability distribution of values.  We are thus lead to
associate a ``fuzzy'' nature to quantum space-time points by this
asymptotic construction. Note thus that the question of whether two
observations $(u,\omega,R;\zeta)$ and $(u',\omega',R';\zeta')$ ``see''
the same point can only be determined probabilistically.

As we just mentioned, there are equivalence classes (topological
2-spheres) of observation points, i.e. points in the 6-dimensional
observation space, which correspond to the same {\it spacetime} point.
In the quantum theory, we could raise the following question: Are there
sets of observation points which are equivalent in the above sense,
i.e. define the ``same'' $\widehat{x}^a$? While we have no conclusive
answers yet, there are possible directions in which to explore this
question. For example, we could consider a collection of observation
points to be ``equivalent'' if the corresponding spacetime point
operators mutually commute. Weaker alternatives would be to look for
sets of $(u,\omega,R;\zeta)$ such that the $\widehat{x}^a =
x^a(u,\omega,R;\zeta; [\widehat{data}])$ possess {\it some} common
eigenstates with the same eigenvalues, or the same expectation values
in some quantum states.  These are only some of the questions that
remain to be thought about and explored.

Finally, the algebraic structure of the ``quantum spacetime'' defined
in this way is characterized by the commutation relations between the
spacetime point operators.  These are given in (\ref{explccr}).  We
suspect that some relevant physical or mathematical result is hidden in
these relations; but we have not been able, so far, to get to a fully
convincing understanding of them. Two ideas may be relevant in this
context.  First, as the classical dynamics of a particle is fully
determined by its gravitational interactions, one is tempted to
speculate that its quantum properties can be derived from quantum
geometry as well, and therefore might be hidden in (\ref{explccr}).
Second, the commutation relations (\ref{explccr}) could be relevant to
the present efforts towards understanding quantum spacetime in terms of
noncommutative geometry \cite{noncommutative}. In that context, the
commutative algebra of smooth functions over the manifolds is replaced
by some noncommutative algebra, but it is difficult to find guidelines
for guessing this noncommutative algebra.  The commutation relations
(\ref{explccr}) define a noncommutative algebra that, {\it if} the
Planck constant goes to zero, is equivalent to the commutative algebra
of smooth functions over the manifold. Notice that this noncommutative
algebraic structure is not assumed here, rather, it is {\it derived~}
from quantum general relativity.   We leave the analysis of these
suggestions for future investigations.

Notice that the picture of quantum gravity presented here is very far
from conventional local quantum field theory where one assumes that
physical points and the spacetime manifold are well defined to start
with.  It is therefore also very far  from any approach to quantum
gravity based on conventional quantum field theoretical ideas.

\acknowledgments

We would like to thank Abhay Ashtekar for helpful discussions.
This research was supported in part by the National Science Foundation
under grants No. PHY 89-04035 and PHY 92-05109.

%------------------------------------------------------------------
\appendix

\section{Ashtekar's asymptotic quantization of the free Maxwell field
and applications}\label{A}

In the main text, we are interested in the asymptotic quantization of
linearized GR. Since the asymptotic phase spaces of GR and the free
Maxwell theory are very similar, in this appendix we describe the
asymptotic quantization of the free Maxwell field. The quantization
follows very closely the usual construction of the Maxwell Fock space
for initial data on a Cauchy surface~\cite{wald84}. Our aim here is to
derive the standard covariant commutation relations between the Maxwell
tensor in the interior at two different spacetime points, from the
commutation relations on the asymptotic fields, which themselves are
represented on a Hilbert space. Our description of linearized GR in the
main text is completely analogous to this. (In fact, in the absence of
IR sectors, we simply make the substitution
$A\leftrightarrow\sigma$ and $E\leftrightarrow N$.)

The material in Appendices \ref{A1} and \ref{A2} is quite well-known
and is simply Ashtekar's asymptotic quantization of the Maxwell field
and GR \cite{ashtekar87}. We present it here for the sake of
completeness.  We differ from \cite{ashtekar87} in one notable detail,
namely the definitions of the distributional field operators
(\ref{Eop}-\ref{daf}).  Finally, in Appendices A3 and A4 we construct
respectively, an integral representation and then the closed form of
the covariant commutation relations for the Maxwell field.

%------------------------------------------------------------------------
\subsection{Phase space and algebra of observables}\label{A1}

Let $\gamma_a$ denote the connection field in the interior of Minkowski
space.  The Maxwell tensor is then obtained by
$F_{ab}=2\nabla_{[a}\gamma_{b]}$. On ${\cal I}^+$, with null generators
$n^a$, we define $A_a:=\pb{\gamma}_a$ as the restriction of $\gamma_a$
to ${\cal I}^+$, and $E_a:=\partial A_a/\partial u =\pounds_{n} A_a$,
the electric field on ${\cal I}^+$.

The space of solutions to Maxwell's equations is a linear phase space
$\Gamma$, and we can introduce as coordinates on $\Gamma$ the electric
fields $E_a(u,\zeta)$ on ${\cal I}^+$. Note that $E_a(u,\zeta)$ is a
gauge invariant quantity, and it is normal to the null generators of
${\cal I}^+$, namely $E_a(u,\zeta)n^a=0$.  Thus $E_a(u,\zeta)$ is
completely defined by the complex scalar $E(u,\zeta)=-m^aE_a(u,\zeta)$;
i.e., $E_a(u,\zeta)=E\bar{m}_a + \bar{E}m_a$.

For the purposes of easing later calculations, let us introduce some
new notation \cite{rst92}.  Let $\alpha,\beta..$ be
infinite dimensional abstract indices on $\Gamma$ which take values in
the continuous set $(u,\zeta)\in{\cal I}^+$. Thus,
 $
   \{ 	(\delta/\delta      E (u,\zeta))^\alpha,
	(\delta/\delta \bar{E}(u,\zeta))^\alpha
   \}
 $
(respectively
  $ \{
	{\rm d}\!{\rm I}_\alpha      E (u,\zeta),
	{\rm d}\!{\rm I}_\alpha \bar{E}(u,\zeta)
   \}
 )$
is a complex vector (covector) coordinate basis on $\Gamma$ (since
$\Gamma$ is a linear space, we do not make a distinction betwen $\Gamma$
and its tangent space at a point).  Thus, for example, a complex scalar
field on ${\cal I}^+$ is a vector $V^\alpha$ in $\Gamma$, with
``components'' $(V(u,\zeta),\bar{V}(u,\zeta))$.  In the index notation
we have introduced, a vector is represented by
 $
	V^\alpha
     =
	\int_{{\cal I}^{^+}}du\, dS^2\,
       (
	       V (u,\zeta)(\delta/\delta      E (u,\zeta))^\alpha
	+ \bar{V}(u,\zeta)(\delta/\delta \bar{E}(u,\zeta))^\alpha
       ).
 $
We follow the abstract index ``summation'' convention, which, in our
case, since the index takes a continuum of values, leads to an
integral. The action of a covector
 $
	W_\alpha
     =
	\int_{{\cal I}^{^+}}du\, dS^2\,
      (        W (u,\zeta) {\rm d}\!{\rm I}_\alpha      E (u,\zeta)
	+ \bar{W}(u,\zeta) {\rm d}\!{\rm I}_\alpha \bar{E}(u,\zeta)
      )
 $
on a vector $V^\alpha$ is given by
  \begin{equation}
	W_\alpha V^\alpha
     =
	\int_{{\cal I}^{^+}}du\, dS^2\,
      (		V (u,\zeta)     W (u,\zeta)
	+  \bar{V}(u,\zeta)\bar{W}(u,\zeta)
	).
  \end{equation}

In this notation, the symplectic structure on the phase space
\cite{ashtekar87} is given by
  \begin{equation}					\label{symp}
	\Omega_{\alpha\beta}
     =
	\frac{1}{2\pi}  	\int
			\int_{{\cal I}^{^+}}\> du \, dS^2\,
					 \> du'\, dS^{2'}\>
	\delta^2(\zeta-\zeta')\Delta(u-u')\>
	{\rm d}\!{\rm I}_\alpha     E (u ,\zeta )
		{\wedge\hskip-6truept \wedge} \hskip1truept
	{\rm d}\!{\rm I}_\beta \bar{E}(u',\zeta'),
  \end{equation}
where $\Delta(u)={\textstyle{1\over2}} \mbox{sgn}(u)$ is the
skew-symmetric antiderivative of $\delta(u)$, so that
$\delta(u)=\partial\Delta(u)/\partial u$.  Note that the symplectic
structure is a {\em constant\/} real two-form on $\Gamma$, and its
action $\Omega(V,W)$ on two vectors $V^\alpha$  and $W^\alpha$ is given
by
  \begin{equation}
	\Omega_{\alpha\beta} V^\alpha W^\beta
      =
	\frac{1}{2\pi}	\int
			\int_{{\cal I}^{^+}} \>    du\, dS^2\,\,
						du'\, dS^{2'}\>
	\delta^2(\zeta-\zeta')\Delta(u-u')\>
    \left(
		V(u ,\zeta )
	  \bar{W}(u',\zeta')-
	  \bar{V}(u',\zeta')
	        W(u,\zeta)
    \right).
  \end{equation}

There are two other naturally defined constant tensors on $\Gamma$
which are useful. Since the electric fields on ${\cal
I}^+$ are orthogonal to the null generators of ${\cal I}^+$, the
(degenerate) metric on ${\cal I}^+$ defines a non-degenerate metric
tensor on $\Gamma$ itself:
  \begin{eqnarray}					\label{metric}
	Q_{\alpha\beta}					&
							   =
							&
	\int\int_{{\cal I}^{^+}}     du \, dS^2\,\,
			         du'\, dS^{2'} \>
	\delta^2(\zeta-\zeta')\,
	\delta  (u-u')\,
   \left(
	{\rm d}\!{\rm I}_\alpha     E (u ,\zeta )\,
	{\rm d}\!{\rm I}_\beta \bar{E}(u',\zeta')
       +
	{\rm d}\!{\rm I}_\alpha\bar{E}(u ,\zeta )\,
	{\rm d}\!{\rm I}_\beta      E (u',\zeta')
   \right)							\cr
							&
							   =
							&
   2	\int_{{\cal I}^{^+}} du\, dS^2\,\>
	{\rm d}\!{\rm I}_{(\alpha}     E (u,\zeta)\,
	{\rm d}\!{\rm I}_{\beta) }\bar{E}(u,\zeta),
  \end{eqnarray}
whose action on two vectors $V^\alpha$ and $W^\alpha$ is
given by
  \begin{equation}
	Q_{\alpha\beta}V^\alpha W^\beta
     =
	\int_{{\cal I}^{^+}} du\, dS^2\,\>
       \left(
	     	     V (u,\zeta)
    		\bar{W}(u,\zeta)
	   +
		\bar{V}(u,\zeta)
    	             W (u,\zeta)
       \right),
  \end{equation}
Next, consider the linear operator corresponding to the $u$ derivative
of fields on ${\cal I}^+$: $\dot{V}(u,\zeta)\equiv\partial
V(u,\zeta)/\partial u$. This is a $(1,1)$ tensor, defined by
  \begin{equation}
	T^\alpha{}_\beta
	V^\beta
    :=
	\dot{V}{}^\alpha
  \equiv
	\int_{{\cal I}^{^+}}du\, dS^2\,
       (
	  \dot{     V} (u,\zeta)(\delta/\delta      E (u,\zeta))^\alpha
	+ \dot{\bar{V}}(u,\zeta)(\delta/\delta \bar{E}(u,\zeta))^\alpha
       ).							\label{t}
  \end{equation}
It is straigtforward to check that the $u$ derivative operator
satisfying (\ref{t}) can be written as
  \begin{equation}\label{dot}
	T^\alpha{}_\beta
      =
	\int\int_{{\cal I}^{^+}}   du \, dS^2\,
		 	       du'\, dS^{2'}	\>
	\,\delta^2(\zeta-\zeta')
	\frac\partial{\partial u}\delta(u-u')
   \left(
    	\left(
		\frac\delta{\delta E(u,\zeta)}
    	\right)^\alpha
    	{\rm d}\!{\rm I}_\beta E(u',\zeta')
      +
    	\left(
		\frac\delta{\delta \bar{E}(u,\zeta)}
    	\right)^\alpha
    	{\rm d}\!{\rm I}_\beta \bar{E}(u',\zeta')
   \right).
  \end{equation}

In relation to the analogy with the linearized NSF
of GR, we are interested in considering the connections as
characteristic free data on ${\cal I}^+$, rather than the electric
fields.  The connections are now determined, with respect to the
electric fields, as the corresponding elements
 $
	A^\alpha
     =
	\int_{{\cal I}^{^+}}du\, dS^2\,
       (
	       A (u,\zeta)(\delta/\delta      E (u,\zeta))^\alpha
	+ \bar{A}(u,\zeta)(\delta/\delta \bar{E}(u,\zeta))^\alpha
       )
 $
of $\Gamma$ such that
  \begin{equation}				\label{e=ta}
	E^\alpha
    =
	T^\alpha{}_\beta
	A^\beta.
  \end{equation}
Defined in this way, the connections are completely determined by a
single complex scalar field $A(u,\zeta)$.  This single complex scalar
is related to the standard real $A_a$ (introduced earlier) by $A=-m^aA_a$
and represents the two degrees of freedom of the Maxwell fields.  In
order to stay away from infra-red sectors,  the remaining component of
$A_a$ is chosen to vanish, namely $A_an^a=0$ (in this gauge, the
Maxwell connection is equivalently represented by either $A_a$ or
$A$).  Note that $T^\alpha{}_\beta$ is degenerate, since it annihilates
fields which do not depend on $u$; thus, it has no unique inverse.
However, the ambiguity in defining $A^\alpha$ by (\ref{e=ta}) is
precisely the remaining gauge freedom, that of an additive field which
depends only on $\zeta$.

The three tensors $\Omega_{\alpha\beta}, Q_{\alpha\beta},
T^\alpha{}_\beta$ on $\Gamma$ are not all independent. In order to
derive the relation between them, note first that the inverse
$\Omega^{\alpha\beta}$ of the symplectic structure, defined by
$\Omega^{\alpha\beta}\Omega_{\beta\gamma} =
1\!\mbox{I}^\alpha{}_\gamma$ is given by
  \begin{equation}					\label{pois}
	\Omega^{\alpha\beta}
     =
	4\pi  	\int
		\int_{{\cal I}^{^+}} du\, dS^2\,
				     du'\, dS^{2'} \>
	     \delta^2(\zeta-\zeta') \frac\partial{\partial u} \delta (u-u') \>
     \left(
	\frac\delta{\delta E(u,\zeta)}
     \right)^{[\alpha}
     \left(
	\frac\delta{\delta \bar{E}(u',\zeta')}
     \right)^{\beta]}
  \end{equation}
and that the inverse of the metric (\ref{metric}) is given by
  \begin{equation}					\label{invmet}
	Q^{\alpha\beta}
     =
	2 \int_{{\cal I}^{^+}}du\, dS^2\,
     \left(
		\frac\delta{\delta      E(u,\zeta)})
     \right)^{(\alpha}
     \left(
		\frac\delta{\delta \bar{E}(u,\zeta)})
     \right)^{\beta)}.
 \end{equation}
Now, combining Eqs. (\ref{dot}), (\ref{pois}) and (\ref{invmet}), a
short calculation shows that
  \begin{equation}			\label{ss=tq}
	\Omega^{\alpha\beta}
      =
	2\pi T^\alpha{}_\gamma Q^{\beta\gamma}.
  \end{equation}
This relationship will be useful later for defining distributional
operators corresponding to the connections.

We now want to construct the Poisson-bracket algebra of elementary
functions on the phase space, which are to be represented in the
quantum theory by quantum operators.  Since the phase space
is a linear space, it will be most convenient to consider the space of
all (sufficiently smooth) linear functions on $\Gamma$, together with
the constant functions. This space can be parametrized in the following
manner. Let $S\subset\Gamma$ be the space
of complex covector test fields on ${\cal I}^+$. Let $V^\alpha\in
S$, and define a function ${\cal F}_V$ on $\Gamma$, whose value,
evaluated at the point $E^\alpha\in\Gamma$ is given by
  \begin{equation}					\label{linfunc}
	{\cal F}_V[E]:=\Omega_{\alpha\beta}E^\alpha V^\beta.
  \end{equation}
This is a linear function on $\Gamma$. Its gradient is
given by $\nabla_\alpha{\cal F}_V=\Omega_{\alpha\beta} V^\beta$.
The Poisson bracket between any two such functions is
  \begin{equation}					\label{elempb}
     \{		{\cal F}_V[E],{\cal F}_W[E]\}
	\equiv
		 \Omega^{\alpha\beta}\nabla_\alpha{\cal F}_V
			             \nabla_\beta {\cal F}_W
	 =
		-\Omega_{\alpha\beta}V^\alpha W^\beta,
  \end{equation}
where $\Omega^{\alpha\beta}$ is the inverse of the symplectic
structure, defined in (\ref{pois}). Since the function on the RHS of
(\ref{elempb}) is independent of $E^\alpha$, the algebra is closed
under Poisson brackets.  This defines the algebra of elementary
classical functions.

{}From the linear functions (\ref{linfunc}), the classical distributional
electric fields can be obtained via
  \begin{equation}					\label{eclass}
	E^\alpha
      =
	-\Omega^{\alpha\beta}\frac\delta{\delta V^\beta}{\cal F}_V[E]
      =
	- 2\pi T^\alpha{}_\gamma
	      Q^{\beta\gamma}
		\frac\delta{\delta V^\beta}{\cal F}_V[E],
  \end{equation}
where we have used (\ref{ss=tq}). Comparing (\ref{eclass})
with (\ref{e=ta}), and making the same gauge choice for the connection
as before, we see that the distributional connection field is given by
  \begin{equation}					\label{aclass}
	A^\alpha
     =
	- 2\pi                Q^{\beta\alpha}
	  \frac\delta{\delta V^\beta}{\cal F}_V[E].
  \end{equation}
{}From (\ref{elempb}) and the definition (\ref{aclass}) of the
classical distributional connection field on ${\cal I}^+$, we can
obtain the fundamental Poisson bracket between two connections:
  \begin{eqnarray}
	\{ A^\alpha,A^\beta \}
    & =&
				4\pi^2 	Q^{\gamma\alpha}
	      				Q^{\delta\beta }
					   \frac\delta{\delta V^\gamma}
					   \frac\delta{\delta W^\delta}
	\{ {\cal F}_V[E],{\cal F}_W[E] \}		\nonumber\\
    & =&
				4\pi^2 	Q^{\gamma\alpha}
	      				Q^{\delta\beta }
	\Omega_{\delta\gamma}				\nonumber\\
    & =&
	-4\pi
		\int
		\int_{{\cal I}^{^+}} du\, dS^2\,
				     du'\, dS^{2'} \>
	     \delta^2(\zeta-\zeta')
	     \Delta  (u-u') 			 \>
     \left(
		\frac\delta{\delta E(u,\zeta)}
     \right)^{[\alpha}
     \left(
		\frac\delta{\delta \bar{E}(u',\zeta')}
     \right)^{\beta]} 				\label{aa1}
  \end{eqnarray}
On the other hand, in terms of components we have
  \begin{eqnarray}
	\{ A^\alpha,A^\beta \}
    & =&
		\int
		\int_{{\cal I}^{^+}} du\, dS^2\,
				     du'\, dS^{2'} \>
	\{ A(u,\zeta),A(u',\zeta') \}
     \left(
		\frac\delta{\delta E(u,\zeta)}
     \right)^{\alpha}
     \left(
		\frac\delta{\delta E(u',\zeta')}
     \right)^{\beta} 				\nonumber\\
						&
						&
  \hspace{1.5cm}  +
	\{ \bar{A}(u,\zeta),\bar{A}(u',\zeta') \}
     \left(
		\frac\delta{\delta \bar{E}(u,\zeta)}
     \right)^{\alpha}
     \left(
		\frac\delta{\delta \bar{E}(u',\zeta')}
     \right)^{\beta} 				\nonumber\\
						&
						&
   \hspace{1.5cm}+2
	\{ A(u,\zeta),\bar{A}(u',\zeta') \}
     \left(
		\frac\delta{\delta E(u,\zeta)}
     \right)^{[\alpha}
     \left(
		\frac\delta{\delta \bar{E}(u',\zeta')}
     \right)^{\beta]} 					\label{aa2}
  \end{eqnarray}
By comparing (\ref{aa1}) and (\ref{aa2}) (or more directly) we obtain
 \begin{mathletters}					\label{A18}
  \begin{equation}
	\{A(u,\zeta),\bar{A}(u',\zeta')\}
     =
	-2{\pi}{}\Delta(u-u')\delta^2(\zeta-\zeta').
  \end{equation}
and
  \begin{equation}
	\{A(u,\zeta), A(u',\zeta')\}
     =
	\{\bar{A}(u,\zeta),\bar{A}(u',\zeta')\}
     =
	0
  \end{equation}
 \end{mathletters}
These are the fundamental distributional Poisson brackets on the data
on ${\cal I}^+$.

Let us summarize what we have done so far. First we have shown that the
linear space of free data of the Maxwell field can be parametrized by
the characteristic data $A(u,\zeta)$ on ${\cal I}^+$. The data
satisfies the Poisson bracket relations (\ref{A18}).
{}From the characteristic data $A(u,\zeta)$, we can obtain the Maxwell
fields in the interior of the spacetime (see subsection
\ref{section:intccr}), and their corresponding Poisson
brackets. {}From the point of view of quantization, the characteristic
data are not convenient elementary observables, since they correspond
to distributions on $\Gamma$ and cannot be directly represented on a
Hilbert space as bounded self-adjoint operators. However, since the
phase space is linear, we introduced the space of linear functionals on
$\Gamma$ in a particularly convenient way, as the space of smeared
electric fields ${\cal F}_V[E]$ (Eq. (\ref{linfunc})). These smeared
fields satisfy the elementary Poisson bracket relations
(\ref{elempb}).  {}From the smeared electric fields ${\cal F}_V[E]$,
the characteristic data $A(u,\zeta)$ can be re-obtained by the
functional derivative with respect to the test fields, via
(\ref{aclass}).

Now in the quantum theory, the elementary operator algebra that one
works with corresponds to the Poisson bracket algebra of the smeared
fields. Following Ashtekar \cite{ashtekar87}, in the next subsection we
construct a representation of this algebra on an asymptotic Fock
space.  We are primarily interested in the distributional connections
on ${\cal I}^+$, and these can be obtained from the smeared electric
field operators via the quantum analog of (\ref{aclass}). The
distributional operators corresponding to the Maxwell fields in the
interior can be constructed from the distributional connections by
analogy with the classical construction (subsection
\ref{section:intccr}).

Hence, to begin with, let us construct the algebra of elementary
operators which we wish to represent in the quantum theory.  We want to
construct operators $\widehat{E}(V)$ corresponding to the classical
functions ${\cal F}_V[E]$. These smeared field operators are defined to
satisfy the standard commutation relations corresponding to the Poisson
brackets (\ref{elempb})
  \begin{equation}					\label{elemcomm}
	[\widehat{E}(V),\widehat{E}(W)]
     =
	i\hbar
		\widehat{ \{
				{\cal F}_V[E],
				{\cal F}_W[E]
			  \}
			 }
     =
	-i\hbar\Omega_{\alpha\beta}V^\alpha W^\beta	\hat{1}.
  \end{equation}
As we noted, we are primarily interested in operator valued
distributions corresponding to the electric fields and the connections
at a point on ${\cal I}^+$. Thus, in analogy with the classical fields
(see (\ref{eclass})), let us define an operator valued distributional
electric field $\widehat{E}^\alpha \equiv (\widehat{E}(u,\zeta),
\widehat{\bar{E}}(u,\zeta))$ by
 \begin{equation}					\label{def}
	\widehat{E}^\alpha
      =
	-\Omega^{\alpha\beta}\frac\delta{\delta V^\beta}\widehat{E}(V).
  \end{equation}
By contracting (\ref{def}) with $\Omega_{\gamma\alpha}V^\gamma$, one
can see that the smeared field operators are obtained from the
distributional operators in the same manner as the linear functions are
smeared with the test fields:
  \begin{equation}					\label{Eop}
	\widehat{E}(V)
      =:
	\Omega_{\alpha\beta}\widehat{E}^\alpha V^\beta.
  \end{equation}
(Compare (\ref{Eop} with (\ref{linfunc})).

Similarly, in analogy with (\ref{aclass}), we define an operator
valued distribution corresponding to the connection fields as follows:
  \begin{equation}					\label{daf}
	\widehat{A}^\alpha
      =
	-2\pi Q^{\beta\alpha}
			\frac\delta{\delta V^\beta}\widehat{E}(V).
  \end{equation}
Using this definition and the commutator (\ref{elemcomm}), we compute the
commutator between the connection operators
  \begin{equation}
		[ \widehat{A}^\alpha,\widehat{A}^\beta ]
     =
				4\pi^2 \,Q^{\gamma\alpha}
	      				Q^{\delta\beta }
					   \frac\delta{\delta V^\gamma}
					   \frac\delta{\delta W^\delta}
	[ \widehat{E}(V), \widehat{E}(W)]
     =
	4\pi^2	i\hbar\widehat{1}	Q^{\gamma\alpha}
	      				Q^{\delta\beta }
	\Omega_{\delta\gamma}.
  \end{equation}
Evaluating the components of $Q^{\gamma\alpha}Q^{\delta\beta}
\Omega_{\delta\gamma}$, as in the classical case, finally
leads to
 \begin{mathletters}					\label{craa'}
 \begin{equation}
	[\widehat{A}(u,\zeta),\widehat{\bar{A}}(u',\zeta')]
     =-2\pi
	i\hbar\Delta(u-u')\delta^2(\zeta-\zeta')\widehat{1}.
  \end{equation}
and
  \begin{equation}
	[\widehat{A}(u,\zeta), \widehat{A}(u',\zeta')]
     =
	[\widehat{\bar{A}}(u,\zeta),\widehat{\bar{A}}(u',\zeta')]
     =
	0
  \end{equation}
 \end{mathletters}
These are the fundamental free-data commutators~\cite{kozameh86}, on
which the commutators for the interior fields are based. They will be
of use in subsection~\ref{section:closedccr}. In the following
subsection we describe the space of states on which the operators act.

%-----------------------------------------------------------------
\subsection{Asymptotic Fock space}		\label{A2}

We are going to construct the (standard) antiholomorphic
representation~\cite{ashtekar87} of the free data on ${\cal I}^+$ for
the free Maxwell field.  Let $S\subset\Gamma$ be the Schwartz
space of complex spin-1 test fields on ${\cal I}^+$. For any
$V(u,\zeta)\in S$, define the Fourier transform
  \begin{equation}
	{\cal V}(\omega)
    =
	\frac1{\sqrt{2\pi}}
			\int_{-\infty}^\infty du\>
	V(u) e^{i\omega u},
  \end{equation}
and the positive frequency part of $V$
  \begin{equation}
	{}^+\!V(u)
      =
	\frac1{\sqrt{2\pi}}
		\int_0^\infty d\omega\>
	{\cal V}(\omega) e^{-i\omega u},
  \end{equation}
where the dependence on $(\zeta,\bar\zeta)$ is understood.  On $S$ (or
$S^+$), the symplectic structure (\ref{symp}) can be expressed as
  \begin{equation}					\label{fsymp}
	\Omega_{\alpha\beta}
      =
	\frac1{2\pi}\int 			d^2S
		    \int_{-\infty}^\infty
						d\omega \>
	\frac1{i\omega} \>
	{\rm d}\!{\rm I}_\alpha    {\cal E}(\omega,\zeta)
	{\wedge\hskip-6truept \wedge} \hskip1truept
	{\rm d}\!{\rm I}_\beta \bar{\cal E}(-\omega,\zeta).
  \end{equation}
On $S^+$, define a Hermitian inner product
  \begin{equation}					\label{ip1}
	\IP{{}^+\!{V}}
	   {{}^+\!{W}}
    :=
	-\frac{i}\hbar\Omega(\overline{{}^+\!{V}},
				       {}^+\!{W})
     =
	\frac1{2\pi\hbar}\int 			d^2S\>
			 \int_0^\infty
					    \frac{d\omega}\omega\>
	\overline{{\cal V}(\omega,\zeta)}			\>
	          {\cal W}(\omega,\zeta).
  \end{equation}
By inspection, this is positive definite. Note that it can be written
in the more familiar form \cite{wald94}:
  \begin{equation}\label{ipwald}
	\IP{{}^+\!{V}}{{}^+\!{W}}
     :=
	\mu(V,W)
      -
	\frac{i}{2\hbar}\Omega(V,W)
  \end{equation}
where $\mu(V,W)=\frac1\hbar Im\,\Omega(\overline{{}^+\!{V}},{}^+\!{W})$
is a real inner product on $S$.

Let us take the Cauchy completion of $S^+$ under this inner product;
denote it ${\cal H}=\overline{S^+}$. As we will see, ${\cal H}$ is the
one-particle H ilbert space. The inner product (\ref{ip1}) defines a
Hermitian metric on ${\cal H}$:
  \begin{equation}					\label{Hmet}
	G_{{\overline\alpha}\alpha}
	\overline{{}^+\!{V}}^{\overline\alpha}
		  {}^+\!{W}^\alpha
     :=
	\IP{{}^+\!{V}}
	   {{}^+\!{W}}.
  \end{equation}
The introduction of this metric will be useful in what follows.

Consider the space
 $
	{\cal F}
      =
	\oplus_{n=1}^\infty
	\otimes_S
		{\cal H}^n\>
	\oplus\comp .
 $
where $\otimes_S$ stands for the symmetric tensor product.
This consists of kets of the form $\ket{{\bf T}}$:
  \begin{equation}
	\ket{{\bf T}}
     =
	\ket{T_0,
	     T_1^{\alpha_1},\cdots
	     T_n^{\alpha_1  \cdots\alpha_n},
			    \cdots},
  \end{equation}
where $T_0\in\comp$ and
$T_n^{\alpha_1\cdots\alpha_n}=T_n^{(\alpha_1\cdots \alpha_n)}\in
\otimes_S {\cal H}^n={\cal H}\otimes_S\cdots (n \mbox{
times})\otimes_S{\cal H}$ is an element of the symmetric tensor product
of $n$ copies of ${\cal H}$.  \footnote{The antiholomorphic
representation can be easily constructed. For example, the one particle
state is represented by
  \begin{equation}
	\psi_{T_1}[Z]
     :=
	\IP{Z}
	   {T}
      =
	G_{{\overline\alpha}\alpha}
	\overline{Z}^{\overline\alpha}
		  T^\alpha,
  \end{equation}
where $Z^\alpha:={}^+\!{E}^\alpha$.}

On this space of states ${\cal F}$ there is the inner product
obtained by extending the inner product on ${\cal H}$:
  \begin{equation}					\label{ipFock}
	\IP{\bf T}
	   {\bf W}
	=
		\overline{T}_0W_0
	+\sum_{n=1}^\infty G_{{\overline\alpha}_1\alpha_1}\cdots
			   G_{{\overline\alpha}_n\alpha_n}
	\overline{T}_n^{{\overline\alpha}_1\cdots{\overline\alpha}_n}
	          W_n^{\alpha_1\cdots\alpha_n}.
  \end{equation}
The Cauchy completion of ${\cal F}$ defines the desired asymptotic Fock
space.

Now that we have the space of states, let us define the creation
and annihilation operators $\hat{c}({}^+\!{V})$ and $\hat{a}({}^+\!{V})$
respectively. Given an element ${}^+\!{V}\in S^+$, define
  \begin{equation}
     \hat{c}({}^+\!{V})\circ\ket{{\bf T}}
   :=
     \ket{
		 0                   				,
	        {}^+\!{V}^{ \alpha}T_0                          ,\cdots,
      \sqrt{n+1}{}^+\!{V}^{(\alpha}T_n^{\alpha_1\cdots\alpha_n)},\cdots
	}
  \end{equation}
and
  \begin{equation}
	\hat{a}({}^+\!{V})
    \circ
	\ket{{\bf T}}
  :=
	\ket{G_{{\overline\alpha}\alpha}
	     \overline{{}^+\!{V}}^{\overline\alpha}
			      T_1^\alpha		  ,\cdots,
   \sqrt{n}
	     G_{{\overline\alpha}\alpha_n}
	     \overline{{}^+\!{V}}^{\overline\alpha}
			      T_n^{\alpha_1\cdots\alpha_n},\cdots}.
\end{equation}

Using (\ref{Hmet}), a straightforward calculation shows that these
operators satisfy the commutation relations
  \begin{equation}					\label{accomm}
	[\hat{a}({}^+\!{V}_1),
	 \hat{c}({}^+\!{V}_2)
	]
   =
	\IP{{}^+\!{V}_1}
	   {{}^+\!{V}_2}	\hat{1},
  \end{equation}
all other commutators vanishing. One can show that these
operators are Hermitian adjoints of each other, i.e.,
  \begin{equation}					\label{hermrel}
	\hat{a}^\dagger({}^+\!{V})
   =
	\hat{c}        ({}^+\!{V}).
  \end{equation}

In this representation, let us define the smeared electric field
operators
  \begin{equation}					\label{efo}
	\widehat{E}(V)
  :=
	\hbar(\hat{c}({}^+\!{V})
      +
	      \hat{a}({}^+\!{V})).
  \end{equation}
{}From the commutator (\ref{accomm}), we easily see that
  \begin{equation}
	[\widehat{E}(V),
	 \widehat{E}(W)
	]
    =
	2i\hbar^2
		Im\IP{{}^+\!{V}}
		     {{}^+\!{W}}\hat{1},
  \end{equation}
where we have used the form of the inner product (\ref{ipwald}).
It follows that the operators we have defined above in
(\ref{efo}) satisfy the desired commutation relations (\ref{elemcomm}).
Furthermore, from the Hermiticity relations (\ref{hermrel}) between the
creation and annihilation operators, we see that the electric field
operators (\ref{efo}) are themselves Hermitian.

Thus, we have constructed a Hermitian representation of the smeared
electric field operators defined in Appendix A1. From these, via
Eq.~(\ref{daf}) we can obtain the distributional connection operators
$\widehat{A}(u,\zeta)$, which satisfy the commutation relations
(\ref{craa'}). Recall that the connections ${A}(u,\zeta)$ on ${\cal
I}^+$ serve as data for the Maxwell fields in the interior of the
spacetime. In the next section, we will use the commutation relations
(\ref{craa'}) between the distributional connection field operators on
${\cal I}^+$ to compute the commutation relations between the field
operators in the interior of the spacetime.

%-----------------------------------------------------------------

\subsection{Integral representations of the covariant commutation
	    relations}		\label{section:intccr}

The fields in the interior of the spacetime can be reconstructed from
knowledge of the fields at ${\cal I}^+$.  The following is a
reconstruction of the Maxwell fields based on the null-surface
formulation of the background Minkowski spacetime~\cite{kkn85}.

In Minkowski space, the intersection of the future lightcone of an
interior point $x^a$ with ${\cal I}^+$ is a topological sphere
$S^2(x^a)$, denoted as the lightcone cut of $x^a$.  In coordinates
$(u,\zeta)$, the lightcone cut of a fixed point $x^a$ is a two-surface
$u=u(x^a,\zeta)$ in ${\cal I}^+$ given by:
  \begin{equation}
	u = Z_0(x^a,\zeta) = x^a           \ell_a(\zeta)
			      = x^a \eta_{ab} \ell^b,
							\label{lccut}
  \end{equation}
where $\ell^b(\zeta)$ is a constant null vector in Minkowski space (see
(\ref{7.a})).  At any fixed point, $\ell^b(\zeta)$ defines the null
cone by varying $\zeta$.  $Z_0$ denotes the $Z$-function for
Minkowski space.

If the (otherwise free) data $A(u,\zeta)$ is restricted to the
lightcone cut of a particular point $x^a$, it defines a function of six
variables denoted $ A_R   (x^a, \zeta)\equiv
A(Z_0(x^a,\zeta),\zeta)$. By giving asymptotic data $A_R$ on a
lightcone cut, the Maxwell field and connection at the interior point
$x^a$ can be found, essentially by differentiation, from the knowledge
of a real non-local superpotential $F(x^a,\zeta)$ which satisfies the
following differential equation on the sphere:
  \begin{equation}
 	 \eth \bar\eth F 			=
	 \eth \bar     A_R   (x^a, \zeta) 		+
     \bar\eth 	       A_R   (x^a, \zeta) 		\equiv
                       D_{_M}(x^a, \zeta)[A].		\label{maxeqn}
  \end{equation}
Regular solutions to this equation from given data can be found in
integral form
  \begin{equation}
	 		       F(x^a, \zeta;\,[A]) 	=
    \int_{S^2} dS^2_\eta\> G_{_M}(\zeta,\eta)
			   D_{_M}(x^a,\eta)[A],		\label{maxfsoln}
  \end{equation}
where $dS^2_\eta=-2i d\eta\wedge d\bar\eta/(1+\eta\bar\eta)^2$ is the
area form for $S^2$, $[A]$ indicates the functional dependence of the
solution on the free data, and $G_{_M}(\zeta,\eta)$ is a known Green's
function~\cite{ikn89}, given by
  \begin{equation}
  	G_{_M}(\zeta,\eta)
      =
	\frac{1}{4\pi}
	\ln
	   \bigg(
	\frac{(    \zeta \!-\!     \eta)(\bar\zeta\!-\!    \bar\eta)}
	     {(1\!+\!\zeta\bar\zeta)            (1\!+\!\eta\bar\eta)}
	   \bigg)
      =
	\frac{1}{4\pi}
	\ln
	   (
		\ell^a(\zeta)\ell_a(\eta)
	    ).						\label{greem}
  \end{equation}
Note that any function of only $x^a$ can be added to (\ref{maxfsoln})
to obtain another solution of (\ref{maxeqn}) with the same datum.  This
gauge freedom of the solutions to (\ref{maxeqn}) is equivalent to
leaving free the $l$=0 term in their spherical-harmonic expansion.  The
Green's function (\ref{greem}) has the property that if $F$ is given by
(\ref{maxfsoln}), then $\int_{S^2} F = 0$, hence $F$ has no $l$=0 term
in an expansion in spherical harmonics.  Eq.~(\ref{maxfsoln}) thus
gives an integral representation of the superpotential $F(x^a,\zeta)$
in the $l$=0 gauge.

In a general gauge, the Maxwell connection $\gamma_a(x^c)$ is related
to $F(x^a,\zeta)$ by
  \begin{equation}
	\ell^a(\zeta)\nabla_a  F(x^c,\zeta)	=
	\ell^a(\zeta)\gamma_a   (x^c).			\label{gamma}
  \end{equation}
By differentiation of (\ref{gamma}) with respect to $\zeta$ and by
algebraic procedures we can reconstruct $\gamma_a(x^c)$, and
$F_{ab}=2\nabla_{[a}\gamma_{b]}$. Explicitly:
  \begin{equation}
	\gamma_a=\gamma_i\lambda^i_a
  \end{equation}
and
  \begin{equation}
	F_{ab}=2\lambda^i_{[b}\nabla_{a]}\gamma_i,
  \end{equation}
where, by definition,
  \begin{eqnarray}
   	 \gamma_{_1}
    &\equiv&
			      \ell^a \nabla_a              F		\\
    	\gamma_{_+}
    &\equiv&
			-        m^a \nabla_a              F
			-     \ell^a \nabla_a \eth         F
      =
		\eth(         \ell^a \nabla_a              F	)	\\
    	\gamma_{_-}
    &\equiv&
			-\bar    m^a \nabla_a              F
			-     \ell^a \nabla_a     \bar\eth F
      =
	    \bar\eth(         \ell^a \nabla_a              F	)	\\
    	\gamma_{_0}
    &\equiv&
			-(n^a-\ell^a)\nabla_a              F
			-        m^a \nabla_a     \bar\eth F
			-     \ell^a \nabla_a \eth\bar\eth F
      =
	\eth\bar\eth(         \ell^a \nabla_a              F	)	\\
  \end{eqnarray}
and $\lambda^i_a(\zeta) \equiv (\ell_a,m_a,\bar{m}_a,n_a-\ell_a)$,
$i$=0,+,-,1. If $F(x^a,\zeta)$ satisfies (\ref{maxeqn}), then
$F_{ab}(x^c)$ is automatically a solution to the Maxwell equations
$\nabla^a F_{ab}=0$.  It is worth noticing that our $l$=0 gauge implies
that the connection $\gamma_a$ is fixed in the Coulomb gauge, being
explicitly given by
  \begin{equation}
	\gamma_a
    =
	\int_{S^2} dS_{\eta}^2 \;
	    \left(
		\dot{\bar{A}}(x^b\ell_b(\eta),\eta) m_a(\eta)
	     +
		\dot{A}(x^b\ell_b(\eta),\eta) \bar{m}_a(\eta)
	    \right).
  \end{equation}
We return to the gauge issue at the end of this subsection.

In the quantization, we define the operators simply by replacing
the classical variables with their quantum versions, i.e.%
\footnote{We assume, as appears to be done for evolution from
Cauchy data, that the operators corresponding to the restrictions of
the connections on ${\cal I}^+$ to the cuts of $x^a$ exist on the Fock
space.},
  \begin{equation}
	 	      \widehat{F}(x^a, \zeta;\,[A]) 	=
    \int_{S^2} dS^2_\eta\> G_{_M}(\zeta,\eta)
			   D_{_M}(x^a,\eta)[\widehat{A}]
							\label{qmaxfsoln}
  \end{equation}

In the following, we find integral representations of the commutation
relations of $\widehat F$, $\widehat\gamma_a$ and $\widehat{F}_{ab}$ at
different values of their arguments. These follow from the fundamental
commutators (\ref{craa'}) for the free data.

Using the notation
  $		[\widehat{F}	,
		 \widehat{F}'	  ]
   \equiv
		[\widehat{F}(x^a ,\zeta ),
	 	 \widehat{F}(x'^a,\zeta')  ]
  $,
from (\ref{qmaxfsoln}) and (\ref{craa'}) we obtain
  \begin{eqnarray}
    		[\widehat{F} , \widehat{F}' ]
      &=& -2
	\int\!\!
	\int_{S^2}\!\!\!
	\Big(
	     \eth_{\eta } G_{_M}(\zeta ,\eta ) \,
	 \bar\eth_{\eta'} G_{_M}(\zeta',\eta')
	+
	 \bar\eth_{\eta } G_{_M}(\zeta ,\eta ) \,
	     \eth_{\eta'} G_{_M}(\zeta',\eta')
	\Big)							\nonumber\\
  && \hspace{2.5cm}
	[\widehat{     A}(x^a \ell_a(\zeta) ,\zeta ),
	 \widehat{\bar A}(x'^a\ell_a(\zeta'),\zeta')]	\;\;
	d^2\!S_{\eta }
	d^2\!S_{\eta'}						\nonumber\\
   &=&
	4\pi i\hbar
	\int_{S^2}
		   \Big(
			\eth_{\eta} G_{_M}(\zeta ,\eta )
		    \bar\eth_{\eta} G_{_M}(\zeta',\eta )
		+
		    \bar\eth_{\eta} G_{_M}(\zeta ,\eta )
			\eth_{\eta}G_{_M}(\zeta',\eta )
		   \Big)
			\Delta (y\!\cdot\!\ell(\eta))		\;
	d^2\!S_\eta						\;\;\;
	 \widehat{1}
								\;\;,
							\label{commf}
  \end{eqnarray}
where $y\!\cdot\!\ell(\eta)\equiv y^a\ell_a(\eta)$ and $y^a\equiv
x^a-x'^a$.  To obtain this result we used the explicit expression of
the Green's function, Eq. (\ref{greem}), and the method for the
evaluation of integrals on the sphere described in~\cite{knp84}.  With
$m_a\equiv \eth \ell_a$ (see the definition of the null tetrad
$(\ell^a,m^a,\bar{m}^a,n^a)$, equation (\ref{tetrad})),  (\ref{commf})
takes the compact form:
 \begin{eqnarray}
		[
	 	 \widehat{F},
	 	 \widehat{F}'
						]
    &=&
	8\pi i\hbar
	\int_{S^2}
		\frac{
			\ell^a(\zeta)
			\ell^b(\zeta')
			   		m_{(a}\bar{m}_{b)}
			}
		      {	\ell(\zeta) \!\cdot\!\ell	\;\;
			\ell(\zeta')\!\cdot\!\ell
		        }
			\Delta (y\!\cdot\!\ell)		\;
	d^2\!S						\;\;\;
		 \widehat{1}				\;\;,
							\label{compf}
  \end{eqnarray}
Here and in the following, we omit the explicit dependence on a dummy
variable, like the integration variable in (\ref{compf}).

In order to find the commutator of $\widehat\gamma_a$, we take two
gradients $\nabla_{\!\!_{\scriptstyle a}}\nabla'_{\! b}$ in the spacetime
arguments of (\ref{compf}) and then contract with
$\ell^a(\zeta)\ell^b(\zeta')$ since
  \begin{equation}
	\ell^a
	\ell'^b \nabla_{\!\!_{\scriptstyle a}}
		\nabla'_{\! b}
	[\widehat{F},\widehat{F}']
   =
	[\ell^a \nabla_{\!\!_{\scriptstyle a}}  \widehat{F},
	 \ell'^b\nabla'_{\! b}                   \widehat{F}']
  =
	[\ell^a  \widehat{\gamma}_{_{\scriptstyle a}}  ,
	 \ell'^b \widehat{\gamma}'_b  ]
  =
	\ell^a
	\ell'^b	 [\widehat{\gamma}_{_{\scriptstyle a}},
		  \widehat{\gamma}'_b]
  								\label{prim}
  \end{equation}
Using (\ref{compf}), we have
  \begin{equation}
	\ell^a
	\ell'^b \nabla_{\!\!_{\scriptstyle a}}
		\nabla'_{\! b} 		[\widehat{F},\widehat{F}']
    =
	-8\pi i \hbar				 		\;
	          \ell^a \ell'^b
	\int_{S^2}
		  m_{(a} \bar{m}_{b)}
				     \dot{\delta}(y\!\cdot\!\ell) \;\,
	d^2\!S							  \;\;\;
	 \widehat{1}
						\label{aux}
  \end{equation}
where we use the notation $\dot{f}(x)\equiv df(x)/dx$. Since the
integral in the right-hand side of (\ref{aux}) is not a function of
$(\zeta,\zeta')$, then, from (\ref{prim}) and (\ref{aux}), it follows
that
  \begin{equation}
	[\widehat{\gamma}_{_{\scriptstyle a}},
	 \widehat{\gamma}'_b]
    =
 	-8\pi i\hbar
	\int_{S^2}
		  m_{(a} \bar{m}_{b)}
				     \dot{\delta}(y\!\cdot\!\ell)  \;\,
	d^2\!S							   \;\;\;
	 \widehat{1}
							\label{commg}
  \end{equation}
This is an integral representation of the commutator of
$\widehat{\gamma}_a$ at two different points, in the interior of the
spacetime. The reason why it does not resemble the standard commutators
for the Maxwell connection is that we have not made the standard gauge
choice, namely the Lorentz gauge, $\nabla^a\gamma_a = 0$. Instead,  by
choosing the superpotential $F$ as (\ref{maxfsoln}) we have picked the
Coulomb gauge, i.e, $\nabla^a\gamma_a = 0$ and $\gamma_{_0} = 0$.
Interestingly, these gauge conditions are consistent with
(\ref{commg}). Namely, if the operators $\widehat{\gamma}_a$ are
constrained by $\nabla^a\widehat{\gamma}_a = 0$ and
$\widehat{\gamma}_{_0} = 0$, then it should also be true that
$[\nabla^a\widehat{\gamma}_{_{\scriptstyle a}} , \widehat{\gamma}'_b ]
= 0$ and $[\widehat{\gamma}_{_0}, \widehat{\gamma}'_b ] = 0$.  By
taking a gradient $\nabla^a$ and observing that $m\!\cdot\!\ell = 0$,
it is straightforward to see that (\ref{commg}) implies
$[\nabla^a\widehat{\gamma}_{_{\scriptstyle a}} , \widehat{\gamma}'_b ]
= 0$, whereas $[\widehat{\gamma}_{_0}, \widehat{\gamma}'_b ] = 0$ holds
trivially, since $m_a$ has a vanishing timelike component (c.f.
equation (\ref{m}).

%-----------------------------------------------------------------
\subsection{Closed form commutators}		\label{section:closedccr}

In this section we evaluate in closed form the integral representation
of the commutator of the non-local potential $F(x^a,\zeta)$,
Eq.~(\ref{commf}).

In the first place, we rewrite the integrand into two terms (by
``flipping'' an $\eth_{\eta}$ derivative from the Green's functions over
to the step function $\Delta$ while keeping the so-called boundary
terms):
  \begin{eqnarray}
	\lefteqn{  \big(
			\eth_{\eta} G  \bar\eth_{\eta} G'	+
			\eth_{\eta} G' \bar\eth_{\eta} G
		   \big)
			\Delta (y\!\cdot\!\ell)
   =}		&&					\nonumber\\
 &&    \eth_{\eta}
	\Big(
		  \big(
			 	   G  \bar\eth_{\eta} G'	+
			 	   G' \bar\eth_{\eta} G		+
			      \bar R				+
			      \bar R'
		   \big)
			\Delta (y\!\cdot\!\ell)
	\Big)
   -
		   \big(
			 	   G  \bar\eth_{\eta} G'	+
			 	   G' \bar\eth_{\eta} G		+
			      \bar R				+
			      \bar R'
		   \big)
				y\!\cdot\! m   		\;
			\delta (y\!\cdot\!\ell)		\label{a2}
  \end{eqnarray}
where $\bar R$ and $\bar R'$ are assumed to satisfy $\eth_{\eta} G =
\bar R$ and $\eth_{\eta} G' = \bar R'$, respectively ($\bar R$ and
$\bar R'$ are not unique). The integration variable on the sphere is
$\eta$.  The other parameters ($y^a,\zeta,\zeta'$), are fixed.  With the
integrand written in this way, the integral in Eq.~(\ref{commf})
splits into two terms:
  \begin{equation}
	[F,F']
      =	-2\pi i\hbar
		\int_{S^2}
  \eth_{\eta}
	\Big(
		  \big(
			 	   G  \bar\eth_{\eta} G'	+
			 	   G' \bar\eth_{\eta} G		+
			      \bar R				+
			      \bar R'
		   \big)
			\Delta (y\!\cdot\!\hat{\ell})
	\Big)
        d\hat{S}^2
    -
	\int_{S^2}
		   \big(
			 	   G  \bar\eth_{\eta} G'	+
			 	   G' \bar\eth_{\eta} G		+
			      \bar R				+
			      \bar R'
		   \big)
				y\!\cdot\!\hat{m   }		\;
			\delta (y\!\cdot\!\hat{\ell})		\;
	d^2\!S						\;\;.
								\label{a3}
  \end{equation}
The first term is a volume integral on the sphere wich can be evaluated
by a method that combines Stokes' theorem and the theorem of residues
for complex variable~\cite{knp84}. The second term in (\ref{a3}) either
vanishes (if $y\!\cdot\!\ell\neq 0$) or is a line integral, since the
integrand has support only on the line defined by $y\!\cdot\!\ell= 0$.
These two distinctions correspond to $y^a$ being timelike or spacelike,
respectively.

For timelike future-pointing $y^a$ the step function
$\Delta(y\!\cdot\!\ell)$ takes the constant value +1/2.  We will first
evaluate (\ref{commf}) in this case, and then extend the result to
timelike past-pointing $y^a$ by simply multiplying by an overall minus
sign.

The commutator (\ref{commf}) is reduced to
  \begin{equation}
	[F,F']    = -\pi  i\hbar
	\int_{S^2}
	\eth_{\eta}  \big(
			 	   G  \bar\eth_{\eta} G'	+
			 	   G' \bar\eth_{\eta} G		+
			      \bar R				+
			      \bar R'
		   \big)
	d^2\!S						\;\;.
								\label{a4}
  \end{equation}
{}From (\ref{greem}) the following are obtained:
  \begin{equation}
   \bar R(\zeta,
	  \eta  ) = 	\frac{1}{4\pi}
			\frac{	\ell(\zeta)\!\cdot\!\ell(\eta)	}
			     {  \ell(\zeta)\!\cdot\!   m(\eta)	}
			\big(
			      \ln     ( \ell(\zeta)\!\cdot\!\ell(\eta)
				       )				-
				1
			\big)					\label{a6}
  \end{equation}
and
  \begin{equation}
	\bar\eth_{\eta} G'		     \equiv
	\bar\eth_{\eta} G(\zeta',
	  		 \eta   ) =
	\frac{1}{4\pi}
	\frac{  \ell(\zeta')\!\cdot\!\bar{m}(\eta)  }
	     {  \ell(\zeta')\!\cdot\!   \ell(\eta)  }\;\;.	\label{a7}
  \end{equation}
Notice that $G$, $\bar R$ and $\bar\eth_{\eta}G'$ are singular at
certain values of $(\eta,\bar\eta)$. This implies that the integral in
(\ref{a4}) must be defined by a limiting process; the integral is
performed on a domain ${\cal D} = S^2 -{\cal B}$ that excludes small
neighborhoods of the singular points, which are eventually shrunk to
zero. By Stokes' theorem, the integral (\ref{a4}) on ${\cal D}$ can be
converted into contour integrations around the singular points.
Furthermore, due to the theorem of residues (with an overall minus
sign), the contour integrals can finally be evaluated by computing the
residues at the simple poles inside ${\cal B}$ (${\cal B}$ consists of
a disjoint union of neighborhoods around singular points)
 \begin{equation}
	[F,F']
      = -2\pi i\hbar
	\oint_{\partial{\cal B}}
		   \big(
			 	   G  \bar\eth_{\eta} G'	+
			 	   G' \bar\eth_{\eta} G		+
			      \bar R				+
			      \bar R'
		   \big)
	\frac{-i    d   \bar\eta}
	     {1\!+\!\eta\bar\eta}
   =
		    8\pi^2i\hbar
	\sum_k \mbox{Res}
		\bigg(
		   \big(
			 	   G  \bar\eth_{\eta} G'	+
			 	   G' \bar\eth_{\eta} G		+
			      \bar R				+
			      \bar R'
		   \big)
			\frac{1}
	     		     {1\!+\!\eta\bar\eta}
		\bigg)\bigg|_{\bar\eta=\bar\eta_k}			\;\;.
								\label{a8}
  \end{equation}
In the evaluation by residues, the variables $(\eta,\bar\eta)$ are
considered independent of each other, the singular points that affect
the integration being those on the variable $\bar\eta$. We are thus
interested in accounting for all the singular points
$\bar\eta=\bar\eta_k$ which are simple poles, while the variable $\eta$
is considered fixed, taking the limiting value $\eta=\eta_k$. Using the
explicit expressions of the scalar products between $\ell^a$ and
$m^a$~\cite{kkn85}
  \begin{eqnarray}
   \ell(\zeta)\!\cdot\!
   \ell(\eta )                &=& \frac{ (    \eta -     \zeta)
				         (\bar\eta - \bar\zeta)  }
				       { (1\!+\!\zeta\bar\zeta)
				         (1\!+\!\eta \bar\eta )}\;,
  								\nonumber\\
    \ell(\zeta)\!\cdot\!
     m   (\eta)               &=& \frac{ (  \bar\eta - \bar\zeta)
				         (1\!+\!\zeta  \bar\eta)  }
				       { (1\!+\!\zeta  \bar\zeta)
				         (1\!+\!\eta   \bar\eta )}\;,
								\label{a9}
  \end{eqnarray}
we see that the integrand in (\ref{a8}) is singular at
$\bar\eta=\bar\zeta,\bar\zeta',-1/\zeta,-1/\zeta'$.  These are simple
poles.  (The apparent pole at $\bar\eta=-1/\eta$ is ignored, since it
does not affect the value of the integral.)  A careful calculation
gives the only non-zero residues
  \begin{equation}
	\mbox{Res}\bigg(
			 	\frac{G  \bar\eth_{\eta} G'}
	     		             {1\!+\!\eta\bar\eta  }
		  \bigg)
			\bigg|_{\bar\eta=\zeta'}		=\;\;\;
	\mbox{Res}\bigg(
			 	\frac{G' \bar\eth_{\eta} G }
	     		             {1\!+\!\eta\bar\eta  }
		  \bigg)
			\bigg|_{\bar\eta=\zeta }		=\;\;
	\frac{1}{4\pi}
	\ln \big(\ell\!\cdot\!\ell')				\label{a10}
  \end{equation}
and
  \begin{equation}
	\mbox{Res}\bigg(
			 	\frac{ \bar R             }
	     		             {1\!+\!\eta\bar\eta  }
		  \bigg)
			\bigg|_{\bar\eta=-1/\zeta }		=\;\;\;
	\mbox{Res}\bigg(
			 	\frac{ \bar R'            }
	     		             {1\!+\!\eta\bar\eta  }
		  \bigg)
			\bigg|_{\bar\eta=-1/\zeta'}		= \;
	\frac{1}{4\pi}
						       \;\;.	\label{a11}
	  \end{equation}
Therefore the commutator for the non-local potential $F$ for
future-pointing timelike separation $y^a$ in closed form is
  \begin{equation}
  	[F,F'] = 2\pi i\hbar
		\left(
			\ln (\ell\!\cdot\!\ell') + 1
		\right)					\;\;.
            							\label{a12}
  \end{equation}
Likewise, the commutator for the non-local potential $F$ for
past-pointing timelike separation $y^a$ in closed form is
  \begin{equation}
	[F,F'] = -2\pi i\hbar
		\left(
			\ln (\ell\!\cdot\!\ell') + 1
		\right)					\;\;.
	    							\label{a13}
  \end{equation}

For spacelike separation $y^a$, the condition $y\!\cdot\!\ell=0$
defines a closed contour on the sphere. This has two immediate
consequences. On one hand, the step function
$\Delta(y\!\cdot\!\hat{\ell})$ changes sign across the contour, which
implies that, in the evaluation by residues, there will be some likely
cancellations, depending on whether the poles are all located on the
same side or are scattered on both sides of the contour.  On the other
hand, there is a non-vanishing contour term that needs to be evaluated
explicitly, in addition to the contribution of the residues.

We will first evaluate (\ref{a3}) for spacelike separation of the form
  \begin{equation}
	y^a = (t, 0, 0, z)	\;\;.				\label{a14}
  \end{equation}
This has the considerable advantage of orienting the contour
$y\!\cdot\!\ell(\eta)=0$ around the $z$-axis; i.e, the contour is a
horizontal circle on the sphere, not necessarily at the equator.  Once
we obtain the result, we will generalize it to an arbitrary spacelike
$y^a$ by means of a general 3-dimensional rotation.

The first term in (\ref{a3}) consists of a combination of the residues
(\ref{a10}) and (\ref{a11}), with appropriate signs depending on
whether the pole is above or below the contour. The step function
$\Delta$ is negative above the contour. The second term in (\ref{a3})
requires a cumbersome calculation, which we outline in the following.

Using standard spherical coordinates $(\theta,\phi)$ on $S^2$, with
$\theta=0$ at the north pole, the stereographic coordinates are given
by $\eta = \cot(\theta/2) e^{i\phi}$, and the condition
$y\!\cdot\!\ell(\eta)=0$ reads
  \begin{equation}
	t-z\cos\hat\theta = 0,
  \end{equation}
defining a circle at a latitude $\theta_{_0}$ given by
$\cos\theta_{_0}=t/z$. The second term in (\ref{a3}) takes the form
  \begin{equation}
   \int_{S^2}
		   \big(
			 	   G  \bar\eth_{\eta} G'	+
			 	   G' \bar\eth_{\eta} G		+
			      \bar R				+
			      \bar R'
		   \big)
				y\!\cdot\!m   		\;
			\delta (y\!\cdot\!\ell)		\;
	d^2\!S
   =
   -
	\int_0^{2\pi}
	   	  \big(
			 	   G  \bar\eth_{\eta} G'	+
			 	   G' \bar\eth_{\eta} G		+
			      \bar R				+
			      \bar R'
		   \big)
		\frac{2\rho_{_0}}{(1\!+\!\rho_{_0}^2)}
		e^{-i\phi}
	d\phi						\;\;,
							\label{a16}
  \end{equation}
where we have used the notation $\rho_{_0} \equiv
\cot(\theta_{_0}/2)$.  Notice that $\rho_{_0}$ increases from 0 at the
south pole to $\infty$ at the north pole, taking the value 1 at the
equator.  The line integral in (\ref{a16}) can be written as the
following contour integral around the unit circle in the complex
plane:
 \begin{equation}
\int_0^{2\pi}
	   	  \big(
			 	   G  \bar\eth_{\eta} G'		+
			 	   G' \bar\eth_{\eta} G		+
			      \bar R				+
			      \bar R'
		   \big)
		\frac{2\rho_{_0}}{(1\!+\!\rho_{_0}^2)}
		e^{-i\phi}
	d\phi
   =
	\oint_{|v|=1}
	   	  \big(
			 	   G  \bar\eth_{\eta} G'		+
			 	   G' \bar\eth_{\eta} G		+
			      \bar R				+
			      \bar R'
		   \big)
		\frac{2\rho_{_0}}{(1\!+\!\rho_{_0}^2)}
	\frac{dv}
	     {iv^2}						\;
   \equiv 							\;
	\mbox{\bf I}					\label{a17}
  \end{equation}
where
  \begin{equation}
	  G  \bar\eth_{\eta} G'
    =	\frac{1}{4\pi}
	  \ln \bigg[
		    \frac{(\rho_{_0} v -     \zeta)
			  (\rho_{_0}/v - \bar\zeta)  }
			 {(1\!+\!\rho_{_0}^2)
			  (1\!+\!\zeta\bar\zeta)     }
	      \bigg]
	  \frac{(1          + \bar\zeta'\rho_{_0} v)}
	       {(\rho_{_0}/v - \bar\zeta'          )}		\label{a18}
  \end{equation}
and
  \begin{equation}
	\bar R
    =
	\frac{1}{4\pi}
	\frac{(    \rho_{_0} v - \zeta   )}
	     {(1 + \rho_{_0}     \zeta/v )}
	\Bigg(
	  \ln \bigg[
 		    \frac{(\rho_{_0} v -     \zeta)
			  (\rho_{_0}/v - \bar\zeta)  }
			 {(1\!+\!\rho_{_0}^2)
			  (1\!+\!\zeta\bar\zeta)     }
	      \bigg]
	   - 1
	\Bigg)						\;\;.\label{a19}
  \end{equation}
With (\ref{a18}) and (\ref{a19}), the contour integral {\bf I} is explicitly
  \begin{eqnarray}
  	\mbox{\bf I}
	=
	\frac{       \rho_{_0}   }
	     {2\pi i(1\!+\!\rho_{_0}^2)}
	  \oint_{|v|=1}
	  \ln \bigg[
		    \frac{(      \rho_{_0} v -     \zeta)
			  (      \rho_{_0}/v - \bar\zeta)  }
			 {(1\!+\!\rho_{_0}^2)
			  (1\!+\!\zeta\bar\zeta)     }
	      \bigg]
	  \frac{(1\!+\!\rho_{_0}^2)(1\!+\!\zeta\bar\zeta')}
	       {(      \rho_{_0}     \!-\!    v\bar\zeta')
		(v\!+\!\rho_{_0}          \zeta          ) }\;	-\;
          \frac{(      \rho_{_0} v \!-\! \zeta )  }
	       {(v\!+\!\rho_{_0}         \zeta ) v} \;\;	\nonumber\\
	+
	  \ln \bigg[
		    \frac{(      \rho_{_0} v -     \zeta')
			  (      \rho_{_0}/v - \bar\zeta')  }
			 {(1\!+\!\rho_{_0}^2)
			  (1\!+\!\zeta'\bar\zeta')     }
	      \bigg]
	  \frac{(1\!+\!\rho_{_0}^2)(1\!+\!\zeta'\bar\zeta)}
	       {(      \rho_{_0}     \!-\!    v \bar\zeta )
		(v\!+\!\rho_{_0}          \zeta'          ) }\;	-\;
          \frac{(      \rho_{_0} v   \!-\!\zeta' )  }
	       {(v\!+\!\rho_{_0}         \zeta' ) v}	 \;\;dv
							\label{I}
  \end{eqnarray}
Technically, the contour integral {\bf I}, Eq. (\ref{I}), can not be
evaluated by residues as it stands, because of the branch cut of the
logarithm at $v=0$. One can rewrite the integrand as a rational
function by introducing a parameter $\tau$ in the argument of the
logarithm, and then differentiating with respect to $\tau$, in the
following fashion:
  \begin{equation}
     \mbox{\bf I} = \mbox{\bf J}(\tau=1)
		  = \int_0^1
			\frac{ d\mbox{\bf J}(\tau)}
			     { d             \tau }
		     d\tau					+
		    \mbox{\bf J}(\tau=0)			\label{a20}
  \end{equation}
where {\bf J}$(\tau)$ is a generalization of {\bf I} defined by
introducing $\tau$, for convenience, as
  \begin{eqnarray}
       \mbox{\bf J}(\tau)				 \equiv
		\frac{             \rho_{_0}   }
		     {2\pi i(1\!+\!\rho_{_0}^2)}
	  \oint_{|v|=1} \!\!\!\!\!\!
	  \ln \bigg[
		    \frac{( \tau \rho_{_0} v \!-\!     \zeta)
			  ( \tau \rho_{_0}/v \!-\! \bar\zeta)  }
			 {(1\!+\!\rho_{_0}^2)
			  (1\!+\!\zeta\bar\zeta)     }
	      \bigg]
	  \frac{(1\!+\!\rho_{_0}^2)(1\!+\!\zeta\bar\zeta')}
	       {(      \rho_{_0}   \!\!-\! v   \bar\zeta')
		(v\!+\!\rho_{_0}          \zeta          ) }	-
          \frac{(      \rho_{_0} v \!-\! \zeta )  }
	       {(v\!+\!\rho_{_0}         \zeta ) v}	\;\;	\nonumber\\
	+
	  \ln \bigg[
		    \frac{( \tau \rho_{_0} v \!-\!     \zeta')
			  ( \tau \rho_{_0}/v \!-\! \bar\zeta')  }
			 {(1\!+\!\rho_{_0}^2)
			  (1\!+\!\zeta'\bar\zeta')     }
	      \bigg]
	  \frac{(1\!+\!\rho_{_0}^2)(1\!+\!\zeta'\bar\zeta)}
	       {(      \rho_{_0} \!\!-\! v      \bar\zeta )
		(v\!+\!\rho_{_0}          \zeta'          ) }	-
          \frac{(      \rho_{_0} v \!-\! \zeta' )  }
	       {(v\!+\!\rho_{_0}         \zeta' ) v}		\;dv
%								\nonumber\\
								\label{a21}
  \end{eqnarray}
Notice that (\ref{a21}) is equal to (\ref{a17}) if $\tau$ is set equal
to 1.  On the other hand, if $\tau$ is set equal to zero then the
$v$-dependence of the logarithm in the integrand of {\bf J} dissapears;
consequently the term {\bf J}$(\tau=0)$ in (\ref{a20}) can be
integrated by residues.

The derivative $d${\bf J}/$d\tau$ is the following:
  \begin{eqnarray}
	\frac{d\mbox{\bf J}}
	     {d\tau}
    =
	\frac{       \rho_{_0}^2  }
	     {2\pi i(1\!+\!\rho_{_0}^2) }
	  \oint_{|v|=1}
	    \bigg(
		   \frac{               v              }
		        {(\tau\rho_{_0} v -   \zeta)   }	+
		   \frac{          1                   }
			{(\tau\rho_{_0}   -\bar\zeta v)}
	    \bigg)
		   \frac{(1\!+\!\rho_{_0}^2)(1\!+\!\zeta\bar\zeta')}
	       		{(      \rho_{_0}   - v         \bar\zeta')
			 (v +   \rho_{_0}          \zeta          )}
	    						\;\;    \nonumber\\
								+
	\bigg(
		   \frac{               v               }
		        {(\tau\rho_{_0} v -   \zeta')   }	+
		   \frac{          1                    }
			{(\tau\rho_{_0}   -\bar\zeta' v)}
	    \bigg)
		   \frac{(1\!+\!\rho_{_0}^2)(1\!+\!\zeta'\bar\zeta)}
	       		{(      \rho_{_0}   - v          \bar\zeta )
			 (v +   \rho_{_0}          \zeta'          ) }
								\;\; dv
								\label{a22}
  \end{eqnarray}
The simple poles that are relevant to the evaluation of $d${\bf
J}/$d\tau$ as a function of $\tau$ are
  \begin{equation}
	v =     \frac{\rho_{_0}    }{\bar\zeta'   }\;,\;\;\;
		\frac{\rho_{_0}    }{\bar\zeta    }\;,\;\;\;
		\frac{\zeta        }{\tau\rho_{_0}}\;,\;\;\;
		\frac{\zeta'       }{\tau\rho_{_0}}\;,\;\;\;
		\frac{\tau\rho_{_0}}{\bar\zeta    }\;,\;\;\;
		\frac{\tau\rho_{_0}}{\bar\zeta'   }\;,\;\;\;
		-         \rho_{_0}      \zeta     \;,\;\;\;
		-         \rho_{_0}      \zeta'    \;.
  \end{equation}
Care must be taken to correctly account for the simple poles that are
inside the unit circle at different values of $\tau$.

After the evaluation by residues, $d${\bf J}/$d\tau$ can be seen to be
an explicit linear combination of terms of the form $1/(a+b\tau)$,
which can be integrated in $\tau$ immediately as a logarithmic
function. The procedure is rather lengthy but entirely
straightforward.

In this way, we have given an outline of the main technical steps
necessary to the evaluation of the second term in (\ref{a3}). By
combining the results obtained separately from the first and second
terms in (\ref{a3}), the following final expression for the commutator
of the non-local potential $F$ at spacelike separation $y^a$ of the
form (\ref{a14}) is obtained, which we present split into four
different cases:
  \begin{itemize}
     \item If $\zeta$ and $\zeta'$ are both above the contour then
  	\begin{equation}
	  [F,F'] = -2\pi i\hbar\left[
			   \ln (\ell\!\cdot\!\ell')    -
			   \ln
				\Big[
			     \frac{(\rho_{_0}^2 - \zeta \bar\zeta')
				   (\rho_{_0}^2 - \zeta'\bar\zeta )}
				  {(1+\rho_{_0}^2)^2
						  \zeta \bar\zeta
						  \zeta'\bar\zeta'}
	   			\Big]			+
			     \frac{(1- \rho_{_0}^2 )}
				  {(1+ \rho_{_0}^2 )}	\right]\;
								\label{a24}
	\end{equation}

     \item If $\zeta$ is above and $\zeta'$ is below the contour then
  	\begin{equation}
	  [F,F'] = -2\pi i\hbar\left[
			   \ln
			    \Big[
				        \frac{(1+\zeta'\bar\zeta')
				         \zeta \bar\zeta }
				     {(1+\zeta \bar\zeta )}
	   		    \Big]				+
			        \frac{(1- \rho_{_0}^2)}
				     {(1+ \rho_{_0}^2)}	\right]\;
								\label{a25}
	\end{equation}

     \item If $\zeta$ is below and $\zeta'$ is above the contour then
  	\begin{equation}
	  [F,F'] = -2\pi i\hbar\left[
			   \ln
			    \Big[
			        \frac{   \zeta'\bar\zeta'
				      (1+\zeta \bar\zeta )}
				     {(1+\zeta'\bar\zeta')}
	   		    \Big]+
			        \frac{(1- \rho_{_0}^2)}
				     {(1+ \rho_{_0}^2)}	\right]\;
								\label{a26}
	\end{equation}

     \item If $\zeta$ and $\zeta'$ are both below the contour then
  	\begin{equation}
	  [F,F'] =  -2\pi i\hbar\left[				-
			   \ln (\ell\!\cdot\!\ell')     	+
			   \ln
				\Big[
			     \frac{(\rho_{_0}^2 - \zeta \bar\zeta')
				   (\rho_{_0}^2 - \zeta'\bar\zeta )}
	   			  {(1+\rho_{_0}^2)^2		   }
	   			\Big]				+
			     \frac{(1- \rho_{_0}^2)}
	   			  {(1+ \rho_{_0}^2)}	\right]\;
	  							\label{a27}
	\end{equation}

  \end{itemize}
The results (\ref{a24}) and  (\ref{a27}) have a regular limit as the
contour is shrunk to zero (unlike (\ref{a25}) and (\ref{a26}), in which
one of the points $\zeta$ or $\zeta'$ would dissappear as the contour
is shrunk to zero).  The contour is shrunk to zero by taking the limits
$\rho_{_0}\rightarrow 0$ (in which case the contour flies off the
sphere at the south pole), and $\rho_{_0}\rightarrow \infty$ (in which
case the contour flies off the sphere at the north pole).  The limiting
values $\rho_{_0}= 0, \infty$ correspond to $t=+z, -z$, i.e., the null
boundaries between the timelike and spacelike regions.  Therefore, it
is expected that (\ref{a24}) has (\ref{a13}) for a limit as
$\rho_{_0}\rightarrow 0$, whereas (\ref{a27}) should have (\ref{a12})
for a limit as $\rho_{_0}\rightarrow \infty$.  This is actually the
case, as can be verified by inspection of (\ref{a24}) and (\ref{a27}).

In order to generalize to an arbitrary spacelike separation $y^a$, in
the following we rewrite the relevant quantities as invariants under
general spatial rotations, keeping the time axis fixed.

We define the unit timelike vector
  \begin{equation}
	T^a \equiv (1,0,0,0)\;,
  \end{equation}
which is invariant under spatial rotations. We also have at our disposal
the vectors $\ell^a$ and $m^a$ given by:
  \begin{eqnarray}
	\ell^a
								  &= &
		 \frac{1}{\sqrt{2}}
		 \bigg( 1					\;,\;
			   \frac{       \zeta + \bar\zeta}
				{ 1\!+\!\zeta   \bar\zeta}	\;,\;
			 -i\frac{       \zeta - \bar\zeta}
				{ 1\!+\!\zeta   \bar\zeta}	\;,\;
			   \frac{-1\!+\!\zeta   \bar\zeta}
				{ 1\!+\!\zeta   \bar\zeta}
		\bigg)						,
							\label{l}\\
	       		m^a
	&\equiv&
		 \eth\ell^a
								  =
		           \frac{1}{\sqrt{2}}
		 \bigg( 0					\;,\;
			   \frac{ 1\!-\!        \bar\zeta^2}
				{ 1\!+\!\zeta   \bar\zeta  }	\;,\;
			 -i\frac{ 1\!+\!        \bar\zeta^2}
				{ 1\!+\!\zeta   \bar\zeta  }	\;,\;
			   \frac{ 2     	\bar\zeta  }
				{ 1\!+\!\zeta   \bar\zeta  }
		\bigg)						.
							\label{m}
  \end{eqnarray}
In terms of these vectors, the relevant quantities in (\ref{a24}),
(\ref{a25}), (\ref{a26}) and (\ref{a27}) take the form
  \begin{eqnarray}
	    t
	&=&
			y \cdot T		  	\nonumber\\
	    z
	&=&
		\sqrt{( y \!\cdot\! T )^2 		-
			y \!\cdot\! y     }	  	\nonumber\\
		   1 + \zeta\bar\zeta
	&=&
		\frac{        2  z               }
		     {           z			-
			         y \!\cdot\! T		+
			\sqrt{2} y \!\cdot\! \ell},	\label{a30}\\
	   \rho_{_0}^2
	&=&
		\frac{  z +	 y \!\cdot\! T	 }
		     {  z -      y \!\cdot\! T   }\;\;,	\nonumber\\
 	  \frac{(\rho_{_0}^2 -\zeta'\bar\zeta )
	 	(\rho_{_0}^2 -\zeta \bar\zeta')}
	       {(1+\rho_{_0}^2)^2	       }
	&=&
	   \frac{    -		   y \!\cdot\! y	\,
			        \ell \!\cdot\! \ell'
		     +	2\,	   y \!\cdot\! \ell	\,
				   y \!\cdot\! \ell'  }
		{(                 z
		     -             y \!\cdot\! T
		     +	\sqrt{2}\, y \!\cdot\! \ell  )
		 (                 z
		     -             y \!\cdot\! T
		     +	\sqrt{2}\, y \!\cdot\! \ell' )}	\;\;,	\nonumber
  \end{eqnarray}
where every scalar product is invariant with respect to spatial
rotations.  By substituting (\ref{a30}) into
(\ref{a24}-\ref{a27}), the commutators are generalized to an arbitrary
spacelike separation $y^a$.

%-----------------------------------------------------------------

\section{Evaluation of the null-surface commutator in the case of
timelike  separation }\label{section:closedzccr}

In this appendix we evaluate (\ref{intzccr}) in closed form for a
special range of the parameters $y^a$. As a first step, however, we
rewrite (\ref{intzccr}) in the following form:
  \begin{equation}
	[\widehat{Z},\widehat{Z}']
    =
	-2\pi i\hbar
	\int_{S^2} \eth_\eta ( V \Delta ) - V \eth_\eta\Delta	\;
	d^2\!S_\eta					\label{rewritez}
  \end{equation}
where $V=V(\eta,\zeta,\zeta')$ is given by
  \begin{eqnarray}
					V
      =
		         \eth_\eta    	G'
	             \bar\eth_\eta^2 	G
	+
		         \eth_\eta   	G
	             \bar\eth_\eta^2  	G'
	-
					G'
	  \eth_\eta  \bar\eth_\eta^2	G
	-
					G
	  \eth_\eta  \bar\eth_\eta^2	G'
	+
					Q'
	  \eth_\eta^2\bar\eth_\eta^2	G
	+
					Q
	  \eth_\eta^2\bar\eth_\eta^2	G'			\nonumber\\
	-
					R'
	  \eth_\eta^3\bar\eth_\eta^2	G
	-
					R
	  \eth_\eta^3\bar\eth_\eta^2	G'
  \end{eqnarray}
and the functions $Q$ and $R$ are (non unique) first and second
primitives of $G$ respectively, in the sense that $G=\eth_\eta Q$ and
$G=\eth_\eta^2 R$. A choice of the functions $Q$ and $R$ is given
in Appendix~\ref{section:01Green's}.

If $y^a$ is timelike and future pointing, then $y\!\cdot\!\ell>0$ and
thus $\Delta(y\!\cdot\!\ell)=+\frac12$, constant on the sphere, whereas
$\eth_\eta\Delta(y\!\cdot\!\ell)=\delta(y\!\cdot\!\ell)
\eth_\eta(y\!\cdot\!\ell)=0$ everywhere on the sphere.  Therefore, for
this range of the parameters $y^a$ the commutator reduces to
  \begin{equation}
	[\widehat{Z},\widehat{Z}']
    =-2\pi i\hbar
	\int_{S^2} \eth_\eta \bigg( \frac12 V \bigg)  \;
	d^2\!S_\eta
  \end{equation}
which can be evaluated by residues (see~\cite{knp84}):
  \begin{equation}
	[\widehat{Z},\widehat{Z}']
    =4\pi \hbar
	\sum_j
	\oint_j
		\frac12 \frac{V}{(1+\eta\bar\eta)}	\;
	d\bar\eta
  =8\pi^2i\hbar	\sum_j
		\mbox{Res} \bigg(\frac12 \frac{V}{(1+\eta\bar\eta)}
			   \bigg)
			   \bigg|_{\bar\eta=\bar\eta_j}		\;\;\;.
  \end{equation}
The poles $\bar\eta_j$ are $\bar\zeta\, , \bar\zeta'\, , (\zeta)^{-1}$ and
$(\zeta')^{-1}$. This can be deduced by inspection of the explicit
expression of $V$ which is obtained from the information about the
Green's function that we give in Appendix~\ref{section:01Green's}.  The
evaluation
of the residues at these poles is straightforward, and gives the final
expression
  \[
		[\widehat{Z},\widehat{Z}']
        =-2\pi i\hbar\left(
		          \ell\!\cdot\ell'
          	     \ln (\ell\!\cdot\ell')
      		- \frac16 \ell\!\cdot\ell'
      		+ \frac13	\right)		\;\;.
  \]

%------------------------------------------------------------------

\section{Properties of the Green's function	}\label{section:01Green's}

The function (\ref{Green's}) gives solutions to the following differential
equation for functions $F$ of spin weight 0 on the sphere, with given
spin-weight-0 source $J$:
  \begin{equation}
	\eth^2\bar\eth^2 F = J				\label{diffeq}
  \end{equation}
One of the properties of this Green's function is that, aside from possible
distributional behaviors at $\zeta=\eta$, it is annihilated by application of
the operation $\eth^4\bar\eth^2$ for all values of $\zeta\neq\eta$:
  \begin{eqnarray}
	\bar\eth_\eta
					G(\zeta,\eta)
     &=&
		\frac{1}{4\pi} \ell(\zeta)\!\cdot\!\bar m(\eta)
			\Big(
		\ln	\big(  \ell(\zeta)\!\cdot\!\ell(\eta)
			\big)
		+ 1
			\Big)
							\;\;,	\nonumber\\
	\bar\eth_\eta^2
					G(\zeta,\eta)
    & =&
	\frac{1}{4\pi}
	\frac{\big(\ell(\zeta)\!\cdot\!\bar m(\eta)\big)^2}
	     {     \ell(\zeta)\!\cdot\!\ell(\eta)		     }
							\;\;,	\nonumber\\
	\eth_\eta\bar\eth_\eta^2
					G(\zeta,\eta)
    & = &
		\frac{1}{4\pi}
			\ell(\zeta)\!\cdot\!\bar m(\eta)
			\bigg(
				\frac{1}
		     		     {\ell(\zeta)\!\cdot\!\ell(\eta)}
				- 3
			\bigg)				\;\;,	\nonumber\\
	\eth_\eta^2\bar\eth_\eta^2
					G(\zeta,\eta)
    & = &
		\frac{1}{2\pi}
		 ( 3 \ell(\zeta)\!\cdot\!\ell(\eta) - 2 )\;\;,	\nonumber\\
	\eth_\eta^3\bar\eth_\eta^2
					G(\zeta,\eta)
     &= &
		\frac{3}{2\pi}
		  \ell(\zeta)\!\cdot\! m(\eta) \;\;,	\nonumber\\
	\eth_\eta^4\bar\eth_\eta^2
					G(\zeta,\eta)
    & = &
		0					\;\;.
  \end{eqnarray}
This property allows for the rewriting of (\ref{intzccr}) in the form
(\ref{rewritez}) in the preceding Appendix.

Another useful property of the Green's function is that, up to free
constants of integration, its primitives $Q_{(n)}$ defined by
$\eth_\eta^n Q_{(n)}=G$ can be found recursively.  In general
  \begin{equation}
		Q_{(n)}(\zeta,\eta)
	=
		\frac{H_{(n)}(\ell(\zeta)\!\cdot\!\ell(\eta))}
		     {\big(\ell(\zeta)\!\cdot\!      m(\eta)\big)^n}
  \end{equation}
where $H_{(n)}(x)$ satisfies
  \begin{equation}
		\frac{d^n H_{(n)}}
 		     {dx^n       }(x)
	    =
		          H_{(0)}(x)
	    =
		\frac{1}{4\pi}
		x \, \ln x
						\qquad
					\mbox{or\hspace{1cm}}
		\frac{d H_{(n)}}
 		     {dx       }(x)
	    =
		 H_{(n-1)}(x)	\;\;.		\label{diffhn}
  \end{equation}
Equation (\ref{diffhn}) can be solved by making the ansatz
  \begin{equation}
		H_{(n)}(x) =  x^{n+1}(C_n\ln x - B_n)	\;\;.
  \end{equation}
By imposing (\ref{diffhn}) we find that the parameters $C_n$ and $B_n$
need to satisfy
  \begin{eqnarray}
		C_{n-1} = (n+1) C_n			\nonumber\\
		B_{n-1} = (n+1) B_n - C_n
  \end{eqnarray}
which are solved by
  \begin{eqnarray}
	C_n = \frac{C_0}{(n+1)!}			\nonumber\\
	B_n = \frac{C_0}{(n+1)!} \sum_{i=0}^{n-1}
	      \frac{1}{(n+1-i)}					\;,
  \end{eqnarray}
where $C_0 = \frac{1}{4\pi}$. In this way, we have found a choice of
the generic primitive of $G$ to any desired order (note that the
primitives are not unique).

Here we show explicitly the first and second primitives:
  \begin{eqnarray}
	Q_{(1)}(\zeta,\eta)
     =
	\frac{1}{8\pi}
	\frac{\big(\ell(\zeta)\!\cdot\!\ell(\eta)\big)^2}
	     {\ell(\zeta)\cdot\!          m(\eta)      }
	\Big(
		\ln \big(\ell(\zeta)\!\cdot\!\ell(\eta)\big)
		-\frac12
	\Big)							\\
	Q_{(2)}(\zeta,\eta)
   =
	\frac{1}{24\pi}
	\frac{\big(\ell(\zeta)\!\cdot\!      \ell(\eta)\big)^3}
	     {\big(\ell(\zeta)\cdot\!           m(\eta)\big)^2}
	\Big(
		\ln \big(\ell(\zeta)\!\cdot\!\ell(\eta)\big)
		-\frac56
	\Big)
  \end{eqnarray}
In Appendix~\ref{section:closedzccr}, we have used the notation $Q
\equiv Q_{(1)}$ and $R \equiv Q_{(2)}$.

A third and essential property of the Green's function can be stated in
terms of the solutions of (\ref{diffeq}).  A solution to (\ref{diffeq})
can be found by
  \begin{equation}
	     F_P = \int_{S^2} G(\zeta\eta) J(\eta) \; d^2\!S_\eta
  \end{equation}
(any other solution can be found by adding to $F_P$ a solution to the
homogeneous equation $\eth^2\bar\eth^2 F = 0$).  It can be shown that
$F_P$ has no $l$=0,1 terms in an expansion in spherical harmonics.
Thus {\em the Green's function (\ref{Green's}) provides a decomposition
of a generic solution into its $l$=0,1 part and its $l\ge 2$ part}.
This third property holds as a consequence of the {\em kernel exclusion
property\/} of the Green's functions for $\eth^n$ acting on
spin-weight-s functions; namely, they yield no spherical harmonics of
order $l\in\{s, \cdots, s+n-1\}$ upon integration on the sphere against a
given source~\cite{ikn89}.

%\bibliography{references}     % run latex, bibtex, latex, latex

\begin{thebibliography}{10}

\bibitem{penrose96}
R. Penrose, Gen. Rel. Grav. {\bf 28},  581  (1996).
%
\bibitem{jmp} For various current ideas on quantum geometry, see the
Special Issue {\it Quantum Geometry and Diffeomorphism Invariant Quantum
Field Theory},  J. Math. Phys. 36 (1995).

\bibitem{fkn95a}
S. Frittelli, C.~N. Kozameh, and E.~T. Newman, J. Math. Phys. {\bf 36},  4975
  (1995).

\bibitem{fkn95b}
S. Frittelli, C.~N. Kozameh, and E.~T. Newman, J. Math. Phys. {\bf
36}(9),  4984 (1995).

\bibitem{fkn95c}
S. Frittelli, C.~N. Kozameh, and E.~T. Newman, J. Math. Phys. {\bf 36},  5005
  (1995).
%
\bibitem{ashtekar87}
A. Ashtekar, {\em Asymptotic quantization} (Bibliopolis, Naples, 1987).

\bibitem{fknrt96}
S. Frittelli, C.~N. Kozameh, E.~T. Newman, C. Rovelli, R.~S.
Tate, Fuzzy spacetime from a null-surface version of {GR}, gr-qc/9603061.

\bibitem{fkn95d}
S. Frittelli, C.~N. Kozameh, and E.~T. Newman, J. Math. Phys. {\bf 36}(11),
6397
  (1995).

\bibitem{np66}
E.~T. Newman and R. Penrose, J. Math. Phys. {\bf 5},  863  (1966).
%
\bibitem{kn83} S. L. Kent and E. T. Newman, J. Math. Phys. {\bf
24}(4), 949 (1983).
%
\bibitem{fn96}
S. Frittelli and E.~T. Newman, {\em Pseudo-Minkowskian coordinates in
  asymptotically flat spacetimes}, in preparation.

\bibitem{noncommutative} A. Chamseddine, A. Connes
hep-th/9606001. J. Madore, ``Fuzzy spacetime'', LPTHE Orsay 96/64. B.,
gr-qc/9607065.
Iochum, D. Kastler, T. Sch\"ucker, hep-th/9607158. S. Doplicher, K.
Fredenhagen, J.E. Roberts, Commun. Math. Phys. {\bf172}, 187 (1995).
  A. Chamseddine, J.
Fr\"olich ``Some elements of Connes' non-commutative geometry, and
spacetime geometry'', ETH Z\"urich Preprint, 1996.

\bibitem{wald84}
R.~M. Wald, {\em General Relativity} (The University of Chicago Press,
Chicago,
  1984).

\bibitem{rst92}
R.~S. Tate,  in {\em Quantum gravity, gravitational radiation and large scale
  structure in the universe} (IUCAA, Pune, 1993).

\bibitem{kozameh86}
An independent derivation of these commutators has been given by
C.~N. Kozameh,  in {\em Lecture Notes in Physics}, edited by A.~O. Barut and
  H.~D. Doebner (Springer-Verlag 261, New York, 1986).
%
\bibitem{wald94} R. M. Wald, {\em Quantum Field Theory in Curved
Spacetime and Black Hole Thermodynamics} (The University of Chicago
Press, Chicago, 1994).
%
\bibitem{kkn85}
S.~L. Kent, C.~N. Kozameh, and E.~T. Newman, J. Math. Phys. {\bf 26},  300
  (1985).

\bibitem{ikn89}
J. Ivancovich, C.~N. Kozameh, and E.~T. Newman, J. Math. Phys. {\bf 30},  45
  (1989).

\bibitem{knp84}
C.~N. Kozameh, E.~T. Newman, and J. Porter, Foundations of Physics {\bf 14},
  1061  (1984).
%
\end{thebibliography}
%\bibliographystyle{prsty}

\end{document}